\documentclass[aps,prb,twocolumn,floatfix,showpacs,superscriptaddress]{revtex4-1}
\usepackage{graphicx}
\usepackage[english]{babel}
\usepackage{amsmath}
\usepackage{amssymb}
\usepackage{times}
\usepackage{tensor}

\newcommand{\be}{\begin{eqnarray}}
\newcommand{\ee}{\end{eqnarray}}

\newcommand{\dc}{c^{\dagger}}
\newcommand{\da}{a^{\dagger}}
\newcommand{\nn}{\nonumber}

\def\ket#1{|#1\rangle}
\def\bra#1{\langle #1 |}
\def\ep#1{\langle #1 \rangle}

\begin{document}

\title{Boundary-induced dynamics in 1D topological systems and memory effects of edge modes}

\author{Yan He}
\affiliation{College of Physical Science and Technology,
Sichuan University, Chengdu, Sichuan 610064, China}
\email{heyan\_ctp@scu.edu.cn}

\author{Chih-Chun Chien}
\affiliation{School of Natural Sciences, University of California, Merced, CA 95343, USA.}
\email{cchien5@ucmerced.edu}
\date{\today }

\begin{abstract}
Dynamics induced by a change of boundary conditions reveals rate-dependent signatures associated with topological properties in one-dimensional Kitaev chain and SSH model. While the perturbation from a change of the boundary propagates into the bulk, the density of topological edge modes in the case of transforming to open boundary condition reaches steady states. The steady-state density depends on the transformation rate of the boundary and serves as an illustration of quantum memory effects in topological systems. Moreover, while a link is physically broken as the boundary condition changes, some correlation functions can remain finite across the broken link and keep a record of the initial condition. By testing those phenomena in the non-topological regimes of the two models, none of the interesting signatures of memory effects can be observed. Our results thus contrast the importance of topological properties in boundary-induced dynamics.
\end{abstract}
\maketitle

\section{Introduction}\label{sec:Intro}
The discovery of topological insulators and other materials exhibiting topological properties in their band structures has opened a growing research field (see Refs.~\onlinecite{Hasan10,Qi11,Bernevig_book,ShenBook,AsbothBook} for a review). Due to the bulk-edge correspondence, a nontrivial topological invariant in the bulk indicates localized edge modes that only appear in the presence of boundaries. Interesting transport properties can then arise from those edge modes. In addition to electronic systems, topological systems have been realized in ultracold atoms in optical potentials \cite{Miyake:2013,Aidelsburger:2013,Jotzu:2014} and photonic and phononic systems \cite{Lu14,Yang15}.

Recently there has been growing interest in studying nonequilibrium behavior of topological systems. For example, dynamics of topological superconductors in one and two dimensions undergoing a global quench of interactions reveals decaying topological quantities~\cite{Sacramento14}, and a Chern insulator experiencing a global interaction quench is expected to exhibit edge currents~\cite{Caio15}. There have also been studies on other topological systems after global quenches of parameters~\cite{Wang15,Zeng15,Sacramento16} or piecewise local quenches~\cite{Grushin15}. Here we explore dynamics of topological models after boundary conditions are changed from periodic to open and vice versa. In the absence of topological properties, different boundary conditions should not cause observable effects in a large system. This is not the case for topological systems possessing edge modes that only emerge if there are boundaries. The emergence and disappearance of edge modes when boundary condition changes is the main theme of this work.

Two paradigmatic one-dimensional (1D) topological models will be implemented to demonstrate interesting dynamics induced by a change of boundary conditions. The Kitaev model \cite{Kitaev} exhibits Majorana-fermion states and describes a topological superconductor. The Su-Schrieffer-Heeger (SSH) model \cite{SSH79} was originally proposed as a model for electronic transport in polyacetylene, and its topological properties are summarized in Refs.~\onlinecite{ShenBook,AsbothBook}. The Zak phase of the SSH model has been measured using cold-atoms \cite{Atala13}. According to the classification of topological insulators \cite{Ryu10}, the Kitaev model belongs to the symmetry class D and the SSH model belongs to the class AIII. Therefore, the Kitaev model has particle-hole symmetry with $p$-wave pairing while the SSH model has sublattice (or chiral) symmetry.

In both models we found the densities of edge modes reach steady-state values after the boundary condition transforms from periodic to open. Steady-state behavior in non-topological systems lacking interaction and dissipation has been reported \cite{Lai16}, and here it is observable in the edge-mode dynamics of topological models. Another interesting feature of the edge mode dynamics studied here is the emergence of memory effects, where the rate of boundary transformation is recorded in the steady-state density of edge modes. Here we emphasize the quantum nature of the memory effects and will call them quantum memory effects. In different models the dependence of the steady-state density on the rate of boundary transformation can be different. Memory effects have been proposed in noninteracting quantum systems possessing interesting properties such as a tunable bound state \cite{Cornean13}, a geometric flat-band \cite{Lai16}, or rate-dependent hysteresis \cite{MekenaHyst}. Here we show that memory effects from pure quantum dynamics should also be observable in edge-mode dynamics of topological models.

Moreover, certain correlations can remain finite across a link that is physically broken after the boundary condition changes if the systems are in the topological regimes. This is because of the initial intra-cell correlations that survives the change of boundary condition. In contrast, the correlations will be shown to simply decay to zero across a broken link if the system is topologically trivial. Therefore, topological models are capable of retaining correlations in dynamics. Since experimental realizations of the Kitaev and SSH models are possible, we will briefly summarize their experimental implications.

The paper is organized as follows. Sec.~\ref{sec:theory} briefly summarizes the Kitaev and SSH models and the formalism for investigating their dynamics, along with suitable initial conditions. Sec.~\ref{sec:result} presents the dynamics of the Kitaev and SSH models after a change of boundary conditions. Evidence of quantum memory effects will be found in the edge modes, and topological effects will be contrasted by the results from the topologically trivial counterparts. Sec.~\ref{sec:conclusion} concludes our study. Details of the two models and their analyses are summarized in the Appendix.

\section{Theoretical background}\label{sec:theory}
\subsection{Kitaev model}\label{sec:Kitaev}
The Hamiltonian of the 1D Kitaev model is given by
\be
H_K&=&\sum_{j=1}^N\Big[-w_j(\dc_j c_{j+1}+\dc_{j+1} c_j)-\mu_j \dc_j c_j+ \nonumber \\
& &\Delta_j c_j c_{j+1}+\Delta_j^* \dc_{j+1} \dc_{j}\Big].
\ee
Here we assume the system is arranged so that the $(N+1)$-th site coincides with the first site. In our study we choose $\mu_i=\mu$ to be a uniform constant. The lattice constant $a$ is taken as the unit of length. We assume that
\be
w_j=\left\{
      \begin{array}{ll}
        w_0, & j\neq N; \\
        w(t), & j=N;
      \end{array}
    \right.\qquad
\Delta_j=\left\{
      \begin{array}{ll}
        \Delta_0, & j\neq N; \\
        \Delta(t), & j=N.
      \end{array}
    \right.
\ee
Here $w_0$ and $\Delta_0$ are constants while $w(t)$ and $\Delta(t)$ are functions of time allowing one to change the boundary conditions. In this work we take the linear form $w(t)=w_0(1-t/t_q)$ and $\Delta(t)=\Delta_0(1-t/t_q)$ for $0\le t\le t_q$ and $w(t)=0$ and $\Delta(t)=0$ for $t>t_q$ to model a transformation from periodic boundary condition (PBC) to open boundary condition (OBC) with a characteristic time $t_q$. Similar considerations also allow the system to transform from OBC to PBC. We note that Ref.~\onlinecite{Fagotti16} has studied global reactions to local changes in spin chains by switching a few links, but the system remains periodic.

It is known \cite{Kitaev,Bernevig_book} that the Kitaev chain is topologically nontrivial when $2|w|>|\mu|$. In this case, an open chain with an even number of sites has two zero-energy modes located at the two ends of the chain. On the other hand, when $2|w|<|\mu|$, the Kitaev chain is topologically trivial and there is no zero-energy edge modes in an even-numbered open chain. Details of the topological property of the Kitaev model are summarized in Appendix~\ref{app:Kitaev}.

\subsubsection{Time evolution and initial condition}
The quantum dynamics can be obtained from the equation of motion in the Heisenberg picture $(\hbar\equiv 1)$
\be
\frac{d c_j}{d t}=-i[c_j,H_K].
\ee
We define the following correlation functions to characterize the dynamics
\be
G_{ij}^{11}=\bra{S_0}c_ic_j\ket{S_0},\quad
G_{ij}^{12}=\bra{S_0}c_i\dc_j\ket{S_0},\\
G_{ij}^{21}=\bra{S_0}\dc_ic_j\ket{S_0},\quad
G_{ij}^{22}=\bra{S_0}\dc_i\dc_j\ket{S_0}.
\ee
Here $\ket{S_0}$ is the initial quantum state, and in the following we will skip the label $S_0$ if there is no ambiguity.
It is more convenient to transform to the Majorana fermion representations. Assuming that $\Delta_j$ is real for any $j$, we introduce the Majorana fermion operators 
$a_{2j-1}=c_j+\dc_j, a_{2j}=-i(c_j-\dc_j)$ and
the Hamiltonian becomes
\begin{eqnarray}\label{eq:HKa}
H_K&=&\frac{i}{2}\sum_{j=1}^N [-\mu_j a_{2j-1}a_{2j}+(\Delta_j+w_j)a_{2j}a_{2j+1}+ \nonumber \\ & &(\Delta_j-w_j)a_{2j-1}a_{2j+2} ].
\end{eqnarray}
The evolution of the equal time correlations $\langle a_{i}a_{j}\rangle$ can be obtained accordingly.
The density on site $i$ and current are
\be
n_i=\ep{\dc_ic_i},\quad I_{i,i+1}=2w_i\mbox{Im}\ep{\dc_i c_{i+1}}.
\ee
Here the current is from site $i$ to site $i+1$. From the details shown in Appendix~\ref{app:Kitaev}, all those physical quantities can be expressed in terms of $\langle a_{i}a_{j}\rangle$.

The equations of motion need to be supplemented with proper initial conditions. In Appendix~\ref{app:Kitaev} we present analytic and numerical methods for obtaining suitable initial quantum states with open or periodic boundary condition. The initial state is half-filled and in the ground state, and then the system evolves accordingly.

\subsubsection{Topological edge mode}
Since the Kitaev model with $2|w|>|\mu|$ is topologically nontrivial, an edge mode emerges as an eigenstate with OBC. The edge mode wavefunction satisfies the eigen-equation, Eq.~\eqref{eq:AB} in the Appendix, with $k=0$ and $E_{k=0}=0$.
Thus,
\be
&&(\phi_0)_j=C_1\Big(\frac{\sqrt{w^2-\Delta^2}}{w+\Delta}\Big)^j\sin (j\theta), \\
&&(\psi_0)_j=C_2\Big(\frac{\sqrt{w^2-\Delta^2}}{w+\Delta}\Big)^{L+1-j}\sin[ (L+1-j)\theta].
\ee
Here $C_{1,2}$ are normalization factors and $\theta=\pi-\tan^{-1}\left(\sqrt{4w^2-\mu^2-4\Delta^2}/\mu \right)$.
The edge-mode creation and annihilation operators can be constructed by using Eq.~\eqref{eq:Kitaev_Majorana} with $k=0$.
Note that in terms of Majorana fermions there are two edge modes located at the two ends of the chain. However, the two Majorana fermions will combine into one ordinary-fermion edge mode. The edge mode density corresponds to the occupation of the edge mode and can be obtain from Eq.~\eqref{eq:Majorana_density} with $k=0$.

\subsection{SSH model}\label{sec:SSH}
The Hamiltonian of the SSH model \cite{SSH79} is given by
\be
H_{S}&=&\sum_{i=1}^N [-w_i(\dc_{A,i}c_{B,i}+\dc_{B,i}c_{A,i})- v_i(\dc_{A,i+1}c_{B,i}+
\nonumber \\
& &\dc_{B,i}c_{A,i+1})+ (\mu_i\dc_{A,i}c_{A,i}-\mu_i\dc_{B,i}c_{B,i})].
\ee
Here $A,B$ label the two sublattices. When $w_i=w$, $v_i=v$, and $\mu_i=0$ for all $i$, the system is topologically non-trivial if $v\neq w$ because the Zak phase \cite{Atala13}, or equivalently the winding number \cite{AsbothBook}, is finite. Moreover, the Zak phase changes signs from $v > w$ to $v < w$. By including additional hopping or onsite terms, Ref.~\onlinecite{Li14} shows rich topological properties of generalized SSH models.

We will consider the last inter-cell hopping coefficient to be time dependent with
\be
v_j=\left\{
      \begin{array}{ll}
        v, & j\neq N. \\
        v(t), & j=N.
      \end{array}
    \right.
\ee
Here we consider $v(t)$ to be a linear transformation with a characteristic time $t_q$. The dynamic equations can be derived from $\frac{d c_j}{d t}=-i[c_j,H_S]$. The exact time evolution of the equal-time correlation function $\langle\dc_{A/B,i}c_{A/B,j}\rangle$ can be obtained and monitored in computer simulations.
The initial state is half-filled and placed in the ground state with the corresponding boundary condition. Details of the initial condition is summarized in Appendix~\ref{app:SSH}.

\subsubsection{Topological edge modes}
For the SSH model with an even number of sites and open boundary condition \cite{ShenBook,AsbothBook}, there can be two edge modes if $w>v$ or no edge mode if $w<v$. Moreover, the edge modes survive in the presence of alternating onsite energies, but their energies are no longer pinned at zero. In the topological regime with alternating onsite energies $\mu$ and $-\mu$, the wavefunctions of the two edge modes expressed in terms of the annihilation operators are given by
\be
&&\psi_1=C_1\sum_n\Big(-\frac w v\Big)^{n-1} c_{A,n}, \nonumber \\
&&\psi_2=C_2\sum_n\Big(-\frac w v\Big)^{N-n} c_{B,n}. \label{edge-SSH}
\ee
They correspond to the eigen-energies $\mu$ and $-\mu$, respectively. Here $C_{1,2}$ are normalization factors.

\section{Result and discussion}\label{sec:result}
\subsection{Boundary-induced dynamics of Kitaev model}
\begin{figure}[h!]
\includegraphics[width=0.45\textwidth,clip]{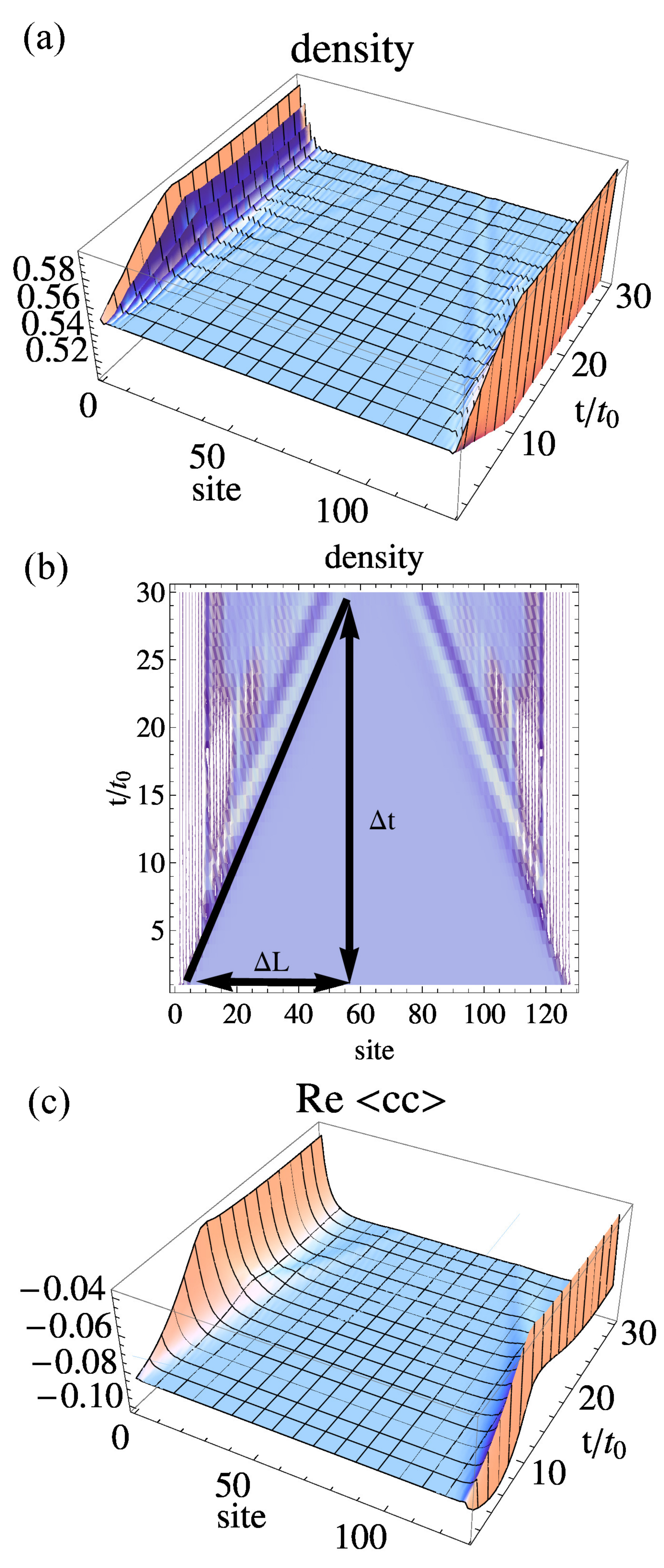}
\caption{Time evolution of the density ((a) for a 3D view and (b) for a top view) and the correlation Re$\ep{c_ic_{i+1}}$ (c) of the 1D Kitaev model as the system transforms from periodic to open boundary condition. The propagation speed is estimated as $\Delta L/\Delta t$. Here $\mu/w=0.3$, $\Delta/w=0.1$, $t_q/t_0=10$, and $N=128$.}
\label{KPToO}
\end{figure}

\begin{figure}[t]
\includegraphics[width=0.45\textwidth,clip]{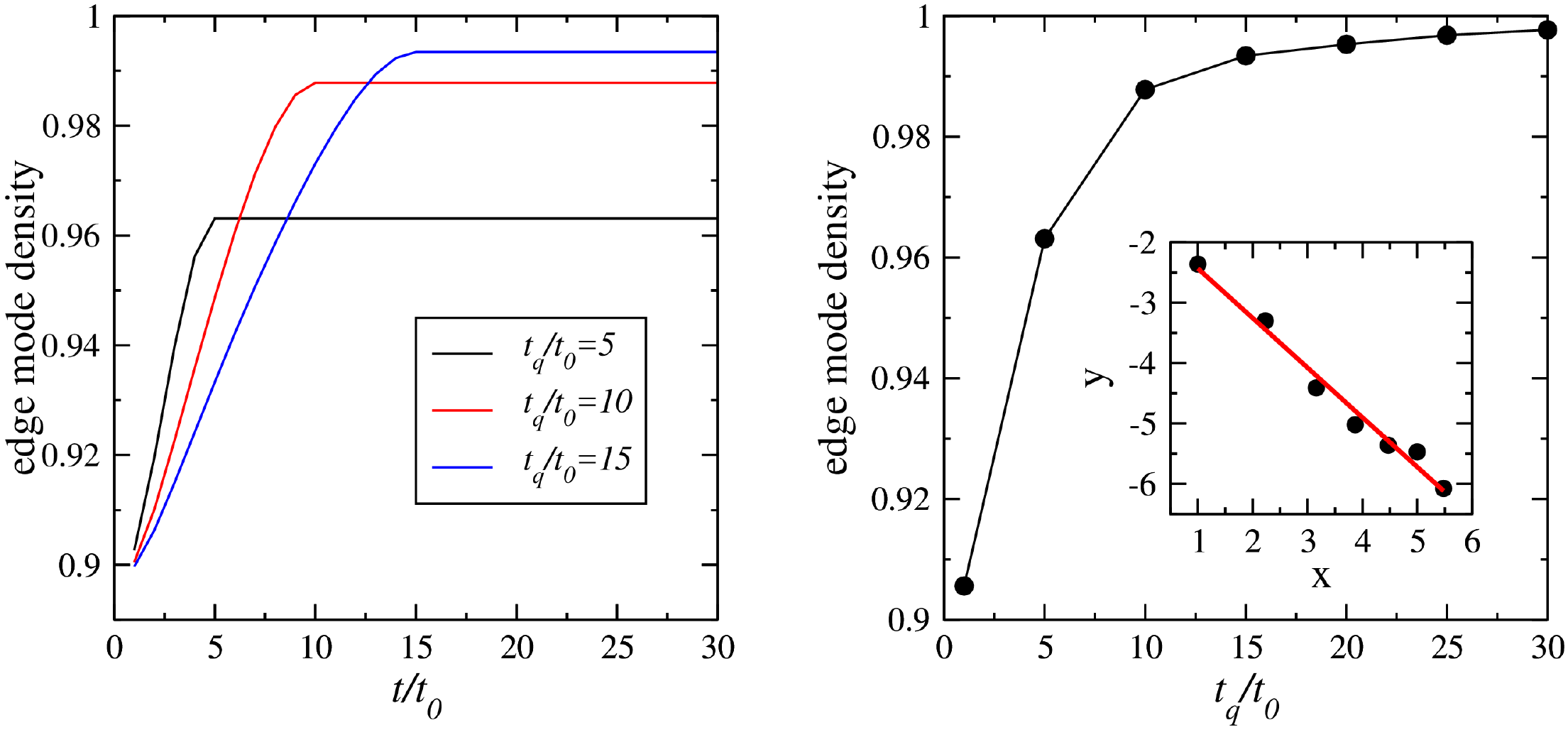}
\caption{The left panel shows the time evolution of the edge mode density for selected ramping times  $t_q/t_0=5,\,10,\,15$, respectively. Here $\mu/w=0.3$, $\Delta/w=0.1$, and $N=128$. The right panel shows the final edge mode density $n_f$ as a function of the ramping time $t_q$. The inset of the right panel plots $y=\ln(1-n_f)$ vs. $x=(t_q/t_0)^{1/2}$ and the red curve is a linear fit.}
\label{edge}
\end{figure}

For the 1D Kitaev model in the topological regime, we choose $\mu/w=0.3$ and $\Delta/w=0.1$, and use $t_0=\hbar/w$ as the time unit. We present the results for $N=128$ sites, and the results are not sensitive to the number of sites if it is reasonably large and even. In Figs. \ref{KPToO} and \ref{edge} we show the results for the time evolution from closed boundary condition to open boundary condition. The transformation of the link between site-$1$ and site-$N$ is assumed to be linear in time with a tunable characteristic time $t_q$.

As shown in Fig.~\ref{KPToO} (a) and (b), the perturbation due to the change of boundary condition propagates with a light-cone structure into the bulk. We found that the propagation speed is close to the maximal group velocity of the Kitaev model. The excitation energy follows the spectrum $E_k=\sqrt{\xi_k^2+4|\Delta|^2\sin^2k}$ with $\xi_k=-2w\cos k-\mu$. Following Refs.~\onlinecite{AshcroftBook,Mekena16}, the group velocity can be found from $v_k=\partial E_k/\partial k$. For the parameters $\mu/w=0.3$, $\Delta/w=0.1$, and lattice constant $a$, the maximal group velocity is $v_{km}\approx 1.87(a/t_0)$, which is very close to the slope of the light cone observed in Figure \ref{KPToO} (b). As $\Delta$ increases, $v_{km}$ decreases. We have verified that the observed light-cone propagation speed of the density and correlation profiles also decreases with increasing $\Delta$, and the value agrees reasonably with the maximal group velocity.

After a transformation from periodic to open boundary condition, two end-points appear and as a consequence, an edge mode should arise at the boundaries. Figure~\ref{edge} shows the rising of the edge mode density. There are two important features. First, the density of the edge mode reaches a steady state exhibiting a plateau after the transformation is completed. The steady state allows us to unambiguously identify memory effects in the growth of the edge mode, which is the second feature. As the ramping time $t_q$ gets longer, the edge mode has higher steady-state density. Such a dependence shows that the steady state of the edge mode is sensitive to the rate of the boundary change, and this is a manifestation of memory effects.

We emphasize that the isolated and noninteracting system considered here is not expected to equilibrate. The edge mode reaches a steady state because it is an eigenstate of the final Hamiltonian with open boundary condition, so its population remains after the boundary transformation is completed. In the right panel of Figure~\ref{edge}, we plot the steady-state value of the edge mode density $n_f$ as function of the ramping time $t_q$.  Within the range we explored, $n_f$ is exponentially approaching $1$ as $t_q$ increases. In the inset, we plot $\ln(1-n_f)$ against $(t_q/t_0)^{1/2}$, and it basically follows a straight line and confirms the exponential dependence.

Here we give pictorial explanations of the steady state and memory effect. A full analytical analysis is hindered by a lack of the full expressions of the eigenstates with open boundary condition. The steady state of the edge mode is because, after the boundary transformation is completed, the edge mode is an exact eigenstate of the final Hamiltonian with open boundary condition. The unitary evolution afterwards thus keeps the occupation of the edge mode intact, so a steady state emerges. We can consider two extreme cases: For a very short ramping time, one would expect, from the sudden approximation, that the the initial state is almost intact and the densities of the final eigenstates are found to be projections of the initial state. Since the edge mode was absent in the initial configuration, the projection leads to a small overlap and this implies a small steady-state density. On the other hand, for a very large ramping time $t_q$, we are practically in the adiabatic limit where $t_q\rightarrow\infty$. Then the initially occupied eigenstate at half filling will smoothly evolve into the edge mode and the final density approaches $1$. For arbitrary finite $t_q$, the resulting $n_f$ will be somewhere in  between those two extremes.

Memory effects usually arise if there are competing time scales. For example, a competition of driving period and relaxation time leads to rate-dependent hysteresis, which is another manifestation of memory effects (see Ref.~\onlinecite{MekenaHyst} and references therein). Here the intrinsic time scale is the hopping time determined by $w$. The edge mode is special in the sense that it is inside the energy gap between the two bands and brings a different energy scale. Filling or emptying the edge mode is at a different time scale compared to the continuum, so different transformation times lead to different final edge-mode densities.

\begin{figure}[t]
\includegraphics[width=0.45\textwidth]{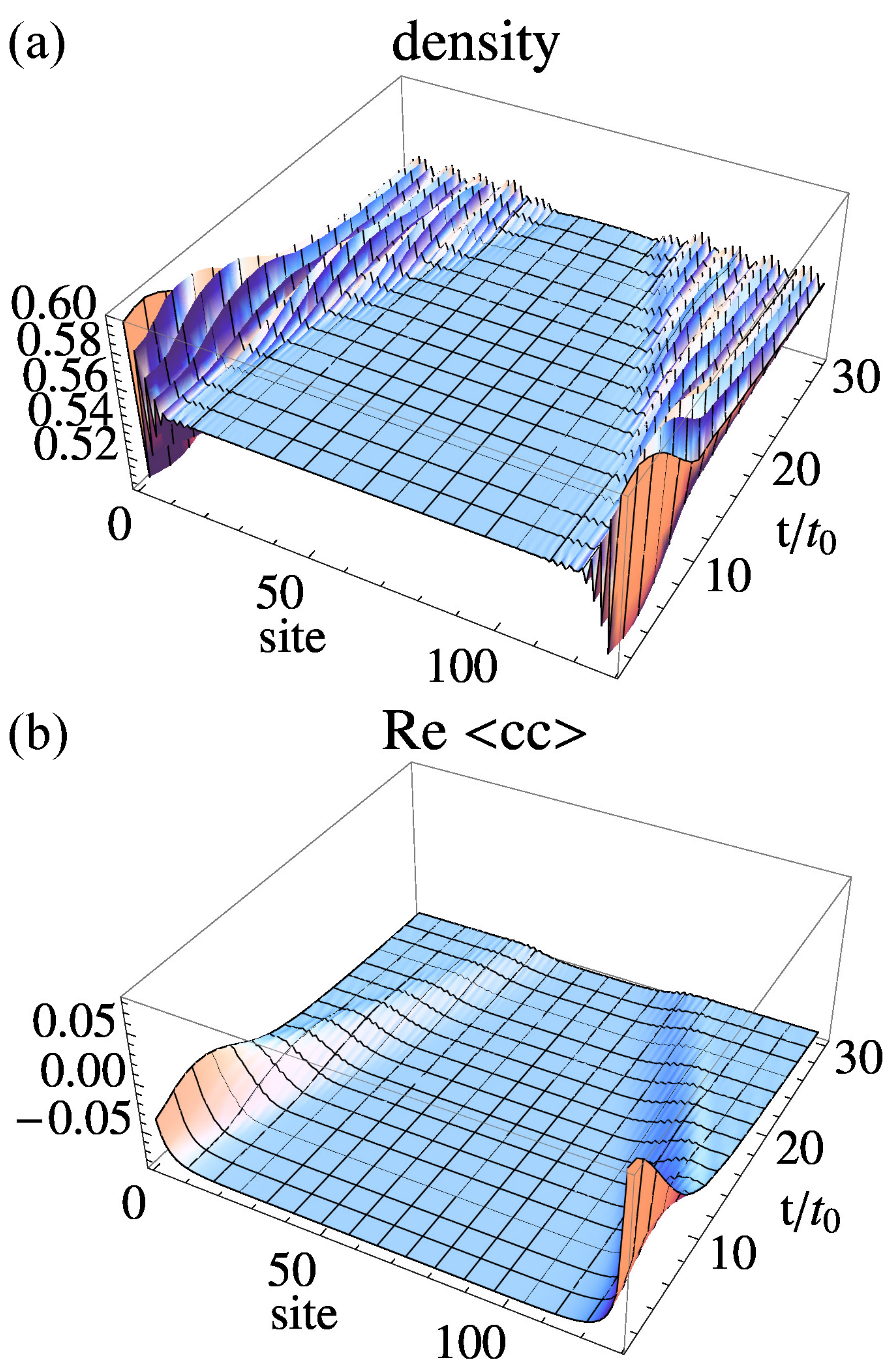}
\caption{Time evolution of (a) density and (b) Re$\ep{c_ic_{i+1}}$ of the 1D Kitaev model from open to periodic boundary condition for $t_q/t_0=10$. Here $\mu/w=0.3$, $\Delta/w=0.1$, and $N=128$.}
\label{KOToP}
\end{figure}
\begin{figure}[t]
\includegraphics[width=0.45\textwidth,clip]{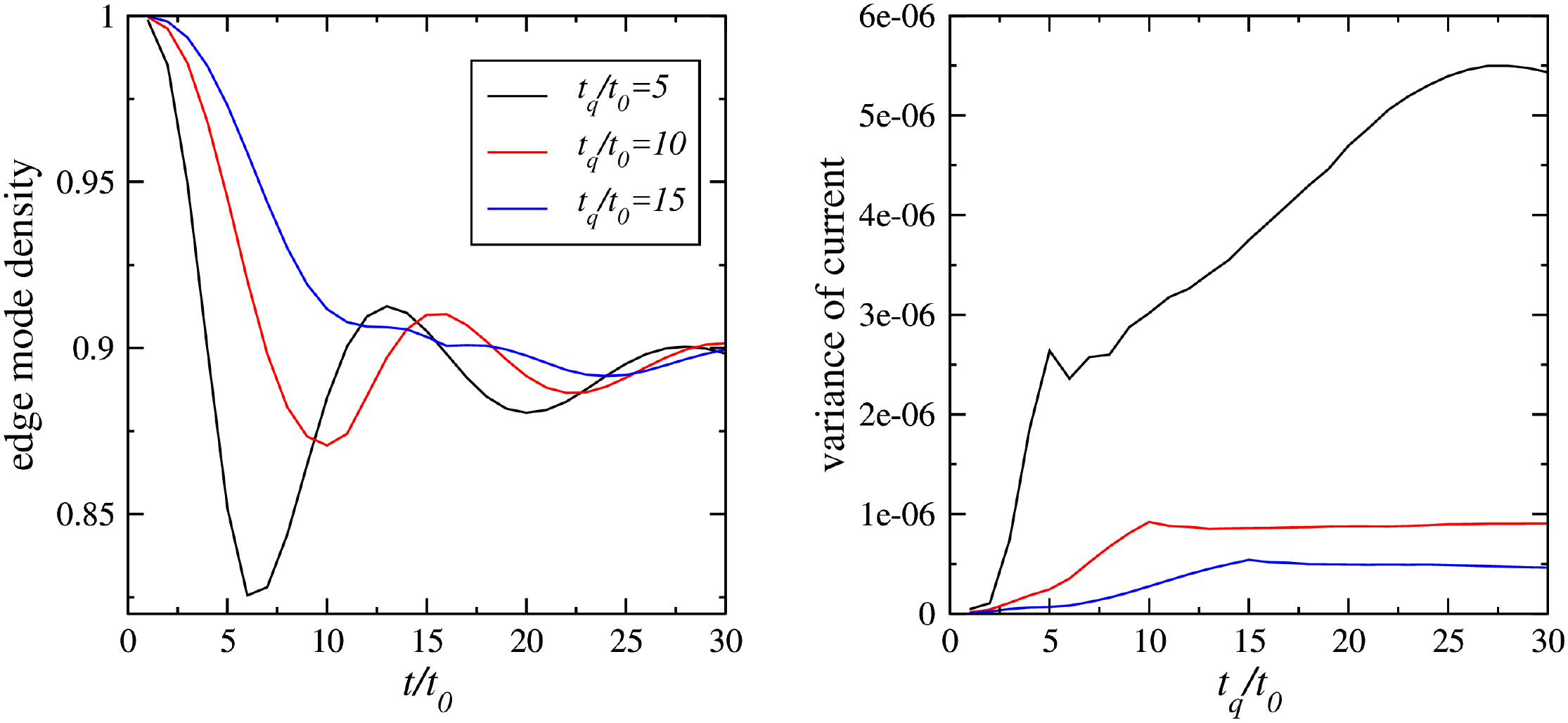}
\caption{The left and right panels show the time evolution of the edge mode density and the variance of the total current (in units of $1/t_0^2)$ of the 1D Kitaev model from open to periodic boundary conditions for $t_q/t_0=5,\,10,\,15$, respectively. Here $\mu/w=0.3$, $\Delta/w=0.1$, and $N=128$.}
\label{edge-r}
\end{figure}

In Figs. \ref{KOToP} and \ref{edge-r}, we show the time evolution from open to periodic boundary conditions. The transformation of the boundary is again assumed to be linear in time with a characteristic time $t_q$. In the initial state with open boundary condition, there is an edge mode and it should decay after the system becomes periodic. Interestingly, the edge mode density exhibits oscillatory behavior as shown in Fig~\ref{edge-r}. The lack of steady-state behavior inhibits a search for memory effects in this case because the edge mode is not an eigenstate of the final Hamiltonian with periodic boundary condition. However, we found that the variance of the total current in the whole system is relatively small and this indicates the system closely follows the averaged behavior.

\begin{figure}[t]
\includegraphics[width=0.45\textwidth]{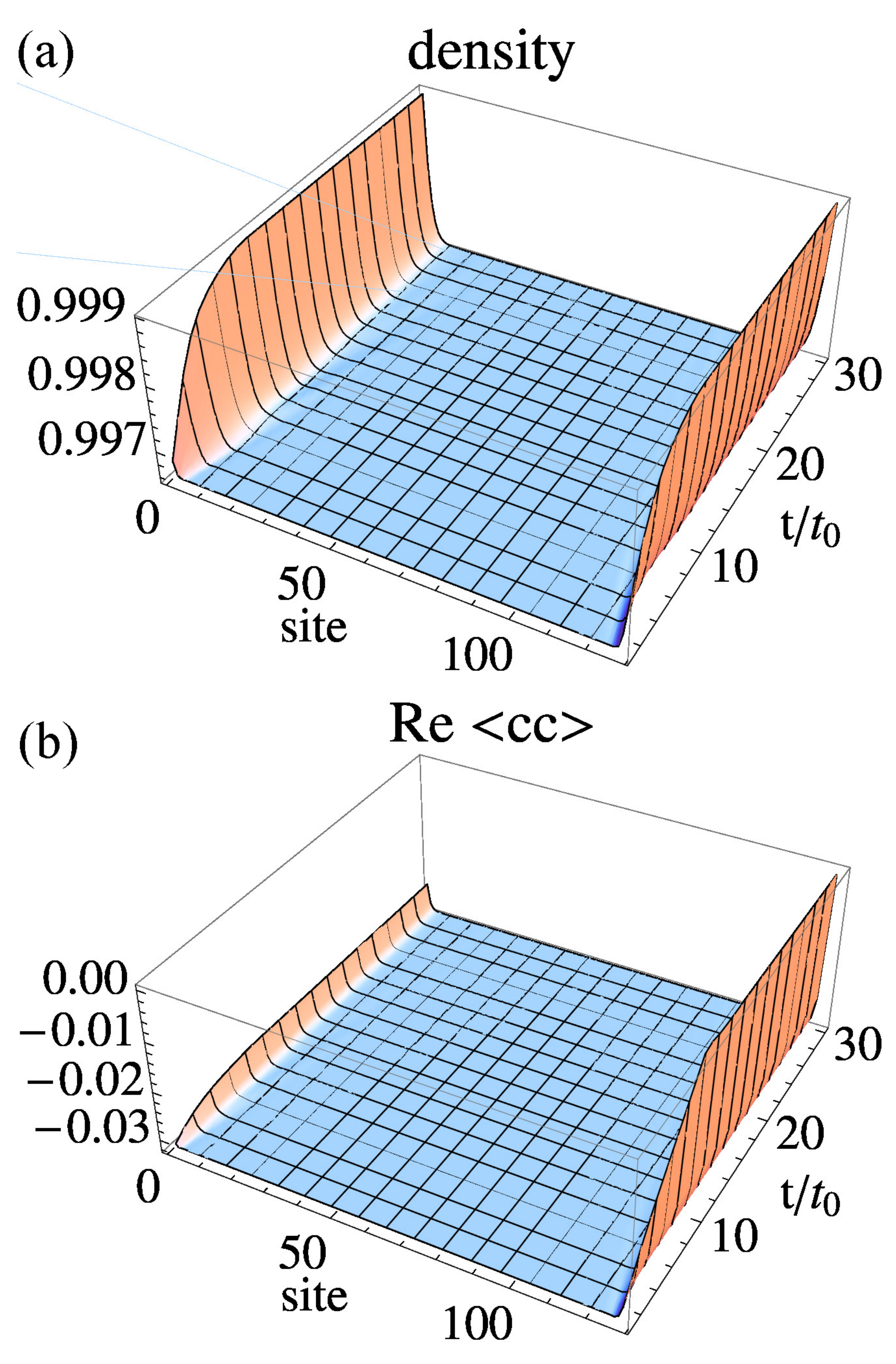}
\caption{The time evolution of (a) density and (b) Re$\ep{c_ic_{i+1}}$ of the Kitaev model from periodic to open boundary conditions for $t_q/t_0=10$. Note the correlation Re$\ep{c_Nc_1}$ decays to zero as shown in (b) (the rising part shown at the rightmost site). The system is in the topologically trivial regime with $\mu/w=2.2$, $\Delta/w=0.1$, and $N=128$.}
\label{Knt}
\end{figure}

To contrast the effects from topological properties, we also present the evolution of the 1D Kitaev model in the topologically trivial regime by taking $\mu/w=2.2$ and $\Delta/w=0.1$. In Figure \ref{Knt} we show the results for a transformation from periodic to open boundary conditions following a linear time dependence with a tunable characteristic time $t_q$.
One important difference between the topological and non-topological cases is the different behavior of the correlation function $\mbox{Re}\langle c_{N}c_1\rangle$ across the broken link. As shown in Fig.~\ref{Knt} (b), in the non-topological regime the correlation $\mbox{Re}\langle c_{N}c_1\rangle$ vanishes completely when the link is broken. In contrast, the correlation decays but remains finite in the topological regime even after the link is physically broken, and one can observe this in Fig.~\ref{KPToO} (c).

One may understand the different behavior of the selected correlation by resorting to the dimer picture of Majorana fermions, where pairs of Majorana fermions form. Firstly,  $\mathrm{Re}\ep{c_Nc_1}=i(\ep{a_{2N}a_1}+\ep{a_{2N-1}a_2})/4$. When we focus on the nearest-neighbor correlation between the Majorana fermions, $\mathrm{Re}\ep{c_Nc_1}$ reveals the correlation between $a_{2N}$ and $a_1$. Next, one may use two chosen sets of parameters to contrast the topological and topologically trivial cases. By choosing $\mu=0$ with finite $w=\Delta$ in the topological regime, the Hamiltonian shown in Eq.~\eqref{eq:HKa} only has Majorana pairs of the type $a_{2j}a_{2j+1}$, which includes the pair $a_{2N}a_{1}$. Therefore, an initially periodic system has correlations between the Majorana fermions on the two side of the link that will be broken after a boundary condition change. In contrast, in the topologically trivial regime one may choose $w=\Delta=0$ with a finite $\mu$. Then, Eq.~\eqref{eq:HKa} only has terms like $a_{2j-1}a_{2j}$, so $a_{2N}a_{1}$ is not directly correlated. For general parameters in the two regime, our numerical simulations confirm that the correlation $\mathrm{Re}\ep{c_Nc_1}$ in the steady state is finite in the topological regime and vanishes in the topologically trivial regime, which is another example of quantum memory effects of initial correlations.

\subsection{Dynamics of SSH model}
For the SSH model, $w$ is the hopping between the A site and B site within one unit cell, and $v$ is the hopping between the B site and A site from two neighboring unit cells. The onsite energies for A and B sites are $\mu$ and $-\mu$, respectively. Here for a chain with even number of sites we take $v/w=1.5$ and $\mu/w=0.1$ in the topological regime and use $t_0=\hbar/w$ as the time unit. We present results for systems with $N=128$ site, and the conclusions are insensitive to the total site number. In contrast to the edge mode of the 1D Kitaev model with open boundary condition, the edge modes of the SSH model are highly oscillatory in space in the initial state.

\begin{figure}[t]
\includegraphics[width=0.45\textwidth]{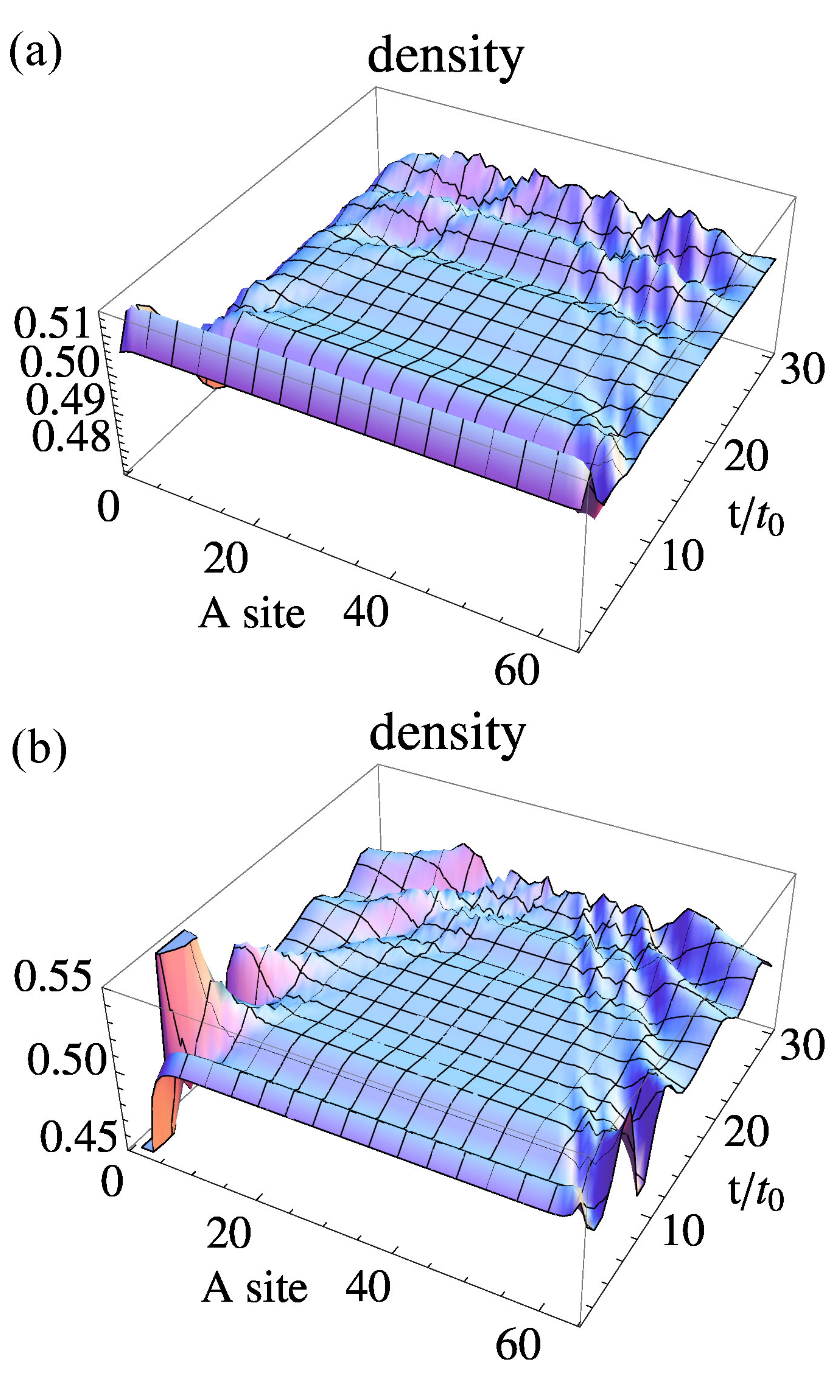}
\caption{Time evolution of the density profile on the A-sublattice for transforming (a) from periodic to open boundary condition and (b) from open to periodic boundary condition. Here   $t_q/t_0=10$, $v/w=1.5$, $\mu/w=0.1$, and $N=128$. The density profile on the B-sublattice has a similar structure.}
\label{A}
\end{figure}

When $\mu=0$, the eigenstates with open boundary condition are close to those with periodic boundary condition. Dynamic signals from a boundary transformation are barely observable on the density profile or correlation functions. In contrast, we found clear dynamic signatures of topological properties with finite values of $\mu$. Due to the alternating $\mu$ on the two sublattices, the density and current distributions for the A and B sites are showing oscillating behavior, so here we show the result on A sites to avoid over-crowded plots. The results for B sites are very similar. In Figure \ref{A} (a), we show the results from periodic to open boundary conditions with a linear transformation of the link connecting site $1$ and site $N$ with a characteristic time $t_q$. Similar to the Kitaev model shown in Fig.~\ref{KPToO} (b), the propagation of the perturbation into the bulk also shows a light-cone structure on the evolution of density profile, and we found that the propagation speed is roughly the maximal group velocity of the system.

Figure \ref{A} (b) shows the evolution of the density profile from open to periodic boundary conditions. In this case the edge modes are initially present, but they decay away in the time evolution. By inspecting the density and current profiles, it is not easy to identify features distinguishing the two types of boundary changes. However, if we focus on the dynamics of the edge modes, visible differences can be observed.

\begin{figure}[t]
\includegraphics[width=0.45\textwidth,clip]{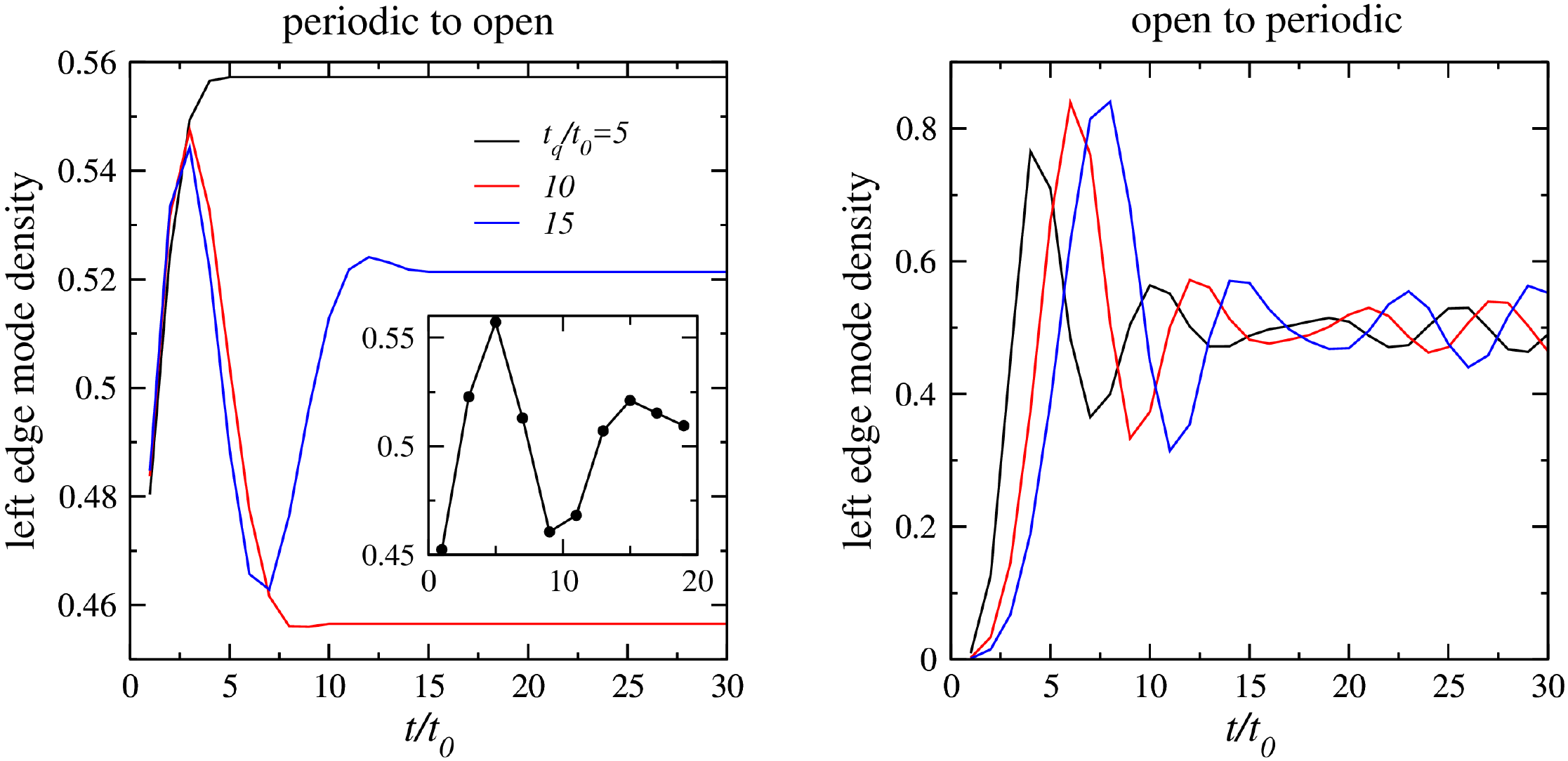}
\caption{Time evolution of the left edge mode density when the system transforms from periodic to open boundary conditions (left panel) and from open to periodic boundary conditions (right panel) for $t_q/t_0=5,\,10,\,15$, respectively. The inset shows the steady-state density (s. s. density) as a function of $t_q$ for the case from periodic to open boundary conditions. Here $v/w=1.5$, $\mu/w=0.1$, and $N=128$.}
\label{AB}
\end{figure}

From Eq.~(\ref{edge-SSH}), the edge mode creation operator can be expressed as a superposition of the fermion operators on different sites. The edge mode density, which reflects the occupation of the edge mode, can be expressed as a superposition of the time-evolved correlation matrix $\langle c^{\dagger}_{i}c_{j}\rangle$. By analyzing the density of the edge mode, we found the edge-mode density exhibits steady-state behavior as shown in Fig.~\ref{AB} for both types of boundary changes. The steady-state values of the edge-mode density allow us to identify memory effects from the boundary-induced dynamics although the oscillations around the steady-state values are smaller in the case transforming to open boundary condition. The steady-state values of the case transforming from periodic to boundary conditions shown in Fig.~\ref{AB} (a) depend explicitly on $t_q$, while the steady-state values of the other transformation shown in Fig.~\ref{AB} (b) are insensitive to $t_q$. The rate-dependent edge-mode density in the former case again serves as evidence that quantum memory effects can be found in topological systems, while there is no memory effect in the latter case.

When compared with the Kitaev model transforming from periodic to open boundary conditions, one can see that the steady-state value of the edge mode density increases monotonically with $t_q$ in the Kitaev model, but the dependence is non-monotonic in the SSH model as one can see in the inset of Fig.~\ref{AB} (a). One may view the alternating onsite energies $\pm\mu$ of the SSH model as an internal bias, which can further tune the dynamics. Such a feature is absent in the Kitaev mode because the onsite energy is uniform. We have checked that the behavior of the edge-mode density of the SSH model depends on $\mu$, and as $\mu\rightarrow 0$ the steady-state values are too small to be discerned.

\begin{figure}[t]
\includegraphics[width=0.42\textwidth,clip]{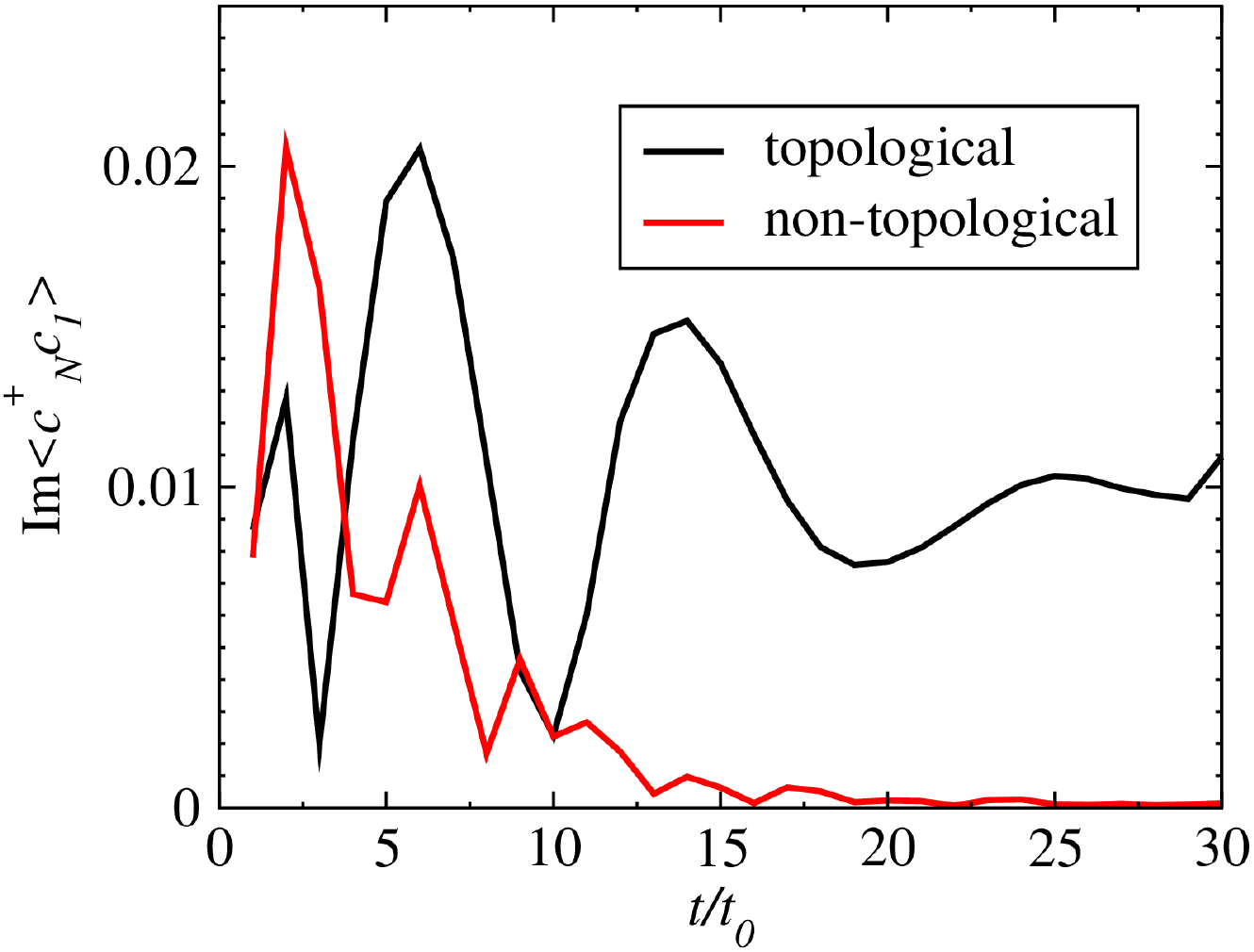}
\caption{Time evolution of Im$\ep{c^{\dagger}_Nc_1}$ of the SSH model from periodic to open boundary conditions for the case with two edge modes (black) and the case with no edge mode (red). The parameters are $v/w=1.5$ ($v/w=0.5$) for the topological (non-topological) case with $\mu/w=0.1$. The topological case retains a finite correlation while the correlation in the non-topological case decays to zero. Here $t_q/t_0=10$ and $N=128$.}
\label{end}
\end{figure}

In Figure \ref{end} we show the time evolution of the correlation function Im$\ep{c^{\dagger}_Nc_1}$ when the system transforms from periodic to open boundary conditions  with two different ratios of $v/w$. An open chain with even number of sites and $v/w>1$ supports two edge modes and is in the topological regime while one with $v/w<1$ does not have any edge mode and is in the non-topological regime. For the former case, we choose $v/w=1.5$ while for the latter case, we choose $v/w=0.5$ in Fig.~\ref{end}. The onsite energies are $\pm\mu/w=\pm 0.1$ for both cases. The link between site $N$ and site $1$ is broken after the transformation, and the correlation Im$\ep{c^{\dagger}_Nc_1}$ for the non-topological case (the case without any edge mode) decays to zero as expected, but the correlation remains finite for the topological case. We have checked the correlation remains the same for much larger system sizes and confirmed the finite correlation is not a finite-size effect.

The finite correlation in the topological regime is again due to the initial inter-cell pair, and it is another manifestation of memory effects in topological systems. One may understand this by analyzing the limit $w=0$ with a finite $v$ in the topological regime. In this case, pairs are localized across each $v$-link and we can focus on the pair that will be separated in the transformation. We may treat the left and right sites of the pair as the two components of a pseudo-spinor. Then the system is similar to the problem of a quantum particle in a double-well potential and the Hamiltonian is $H=\mu\sigma_{z}+v(t)\sigma_{x}$, where $\sigma_{x,y,z}$ are the Pauli matrices. The solution to $i\partial_t \chi=H\chi$ is formally written as $\chi(t)=\exp(-i\mathcal{T}\int^{t}H(t')dt')\chi(0)$, where $\chi$ is a two-component spinor and $\mathcal{T}$ denotes time-ordering. Then $\mbox{Im}\langle c^{\dagger}_{1}c_{N}\rangle=\chi^{\dagger}(t)i\sigma_y\chi(t)$ is associated with the accumulated  dynamic phase when $v(t)$ is reduced to zero. In contrast, in the other limit $v=0$ with a finite $w$ in the non-topological regime, the pairs are across those $w$-links. In this case cutting a $v$ link has no effects on $\mbox{Im}\langle c^{\dagger}_{1}c_{N}\rangle$ and it should remain zero after the transformation. Our numerical results confirm the emergence of memory effect in the selected correlation function.

The observation of quantum memory effects in topological models provides another example of how noninteracting quantum systems can support interesting dynamic phenomena. Quantum memory effects, however, are not unique to topological systems. When compared to previous studies utilizing a tunable bound state~\cite{Cornean13}, a flat-band of dispersionless states~\cite{Lai16}, a competition between the characteristic times of driving and dissipation~\cite{MekenaHyst}, the memory effects in the topological models discussed here offer an additional realization for retaining information of dynamics. 

\subsection{Experimental implications}
The Kitaev model has inspired experimental searches of Majorana fermions in condensed matter systems. For example, superconductor-semiconductor nanowire devices \cite{Mourik12} and superconductor-ferromagnetic chain systems \cite{Perge14} are promising systems for realizing the model. Since those solid-state devices can be engineered, we envision a ring-shape device with a tunable link may be fabricated for simulating the phenomena discussed here.

The SSH model, on the other hand, has been realized using cold-atoms and optical superlattices~\cite{Atala13}. However, cold-atom systems are usually tubes of atomic clouds and may not be suitable for investigating properties with periodic boundary condition. By using a rapid painting potential, ring-shape optical lattices have been realized~\cite{Henderson09}. Future modifications of the painting-potential technique may allow the dynamics discussed here to be simulated. As discussed in Ref.~\onlinecite{ChienNP}, direct observations of edge modes in cold-atom systems can be challenging, but techniques like the quantum-gas microscope \cite{Bakr09} may help resolve the density distribution.

Since correlation functions such as $\mbox{Re}\langle c_{j}c_{j+1}\rangle$ of the Kitaev model or $\mbox{Im}\langle c^{\dagger}_{j}c_{j+1}\rangle$ of the SSH model are useful in identifying memory effects, measurements of those correlations are highly desired but they can be challenging as well. One possible way is to measure the time evolution of the density correlation $\langle n_{j}n_{j+1}\rangle - \langle n_j\rangle\langle n_{j+1}\rangle$. From Wick decomposition \cite{Walecka}, pairwise correlations such as $\langle c^{\dagger}_{j}c_{j}\rangle$ and $\langle c_{j+1}c_{j}\rangle$ will be included in $\langle n_{j}n_{j+1}\rangle$, and the dominant density-density contribution $\langle n_j\rangle\langle n_{j+1}\rangle$ is subtracted. The correlations without memory effects contribute to a background while the correlations with memory effects may exhibit observable rate dependence.

Therefore, boundary-induced dynamics in topological models may still be testable in future experiments with quantum technologies. We caution that dissipation from the environment will eventually relax the system into thermal equilibrium and wash out any dynamic signature in the long-time limit. Measurement of the steady state and memory effect should be performed immediately after the boundary transformation.

\section{Conclusion}\label{sec:conclusion}
We have shown, by tuning only one link in 1D topological models like the Kitaev chain or the SSH model to change boundary conditions, interesting dynamical phenomena emerge. The edge modes exhibit steady-state behavior and allow us to identify quantum memory effects in topological systems when the boundary condition transforms from periodic to open. Correlations across the link broken in the transformation can remain finite in the topological regimes and vanish in topologically trivial regimes. The boundary-induced dynamics thus complements global quench dynamics and reveals non-equilibrium behavior due to topological properties.

\textit{Acknowledgment:} Y. H. thanks the support of NSFC via Grant No. 11404228. C. C. C. thanks the Hellman Family Foundation for partially supporting this work.

\appendix
\section{Details of Kitaev model}\label{app:Kitaev}
This topological property of the Kitaev model can be characterized by a bulk topological invariant called the Majorana number, which is defined by Kitaev in his original paper \cite{Kitaev}. For a one-dimensional $n$-band non-interacting Kitaev chain, the Hamiltonian can be expressed in terms of the Majorana fermions as
$H=\frac{i}{4}\sum_{ab}B_{ab}(k)c_{k,a}c_{k,b}$. Here $c_{k,a}$ and $c_{k,b}$ are annihilation operators of Majorana fermions with $k$ denoting momentum and $a,b=1,\cdots 2n$ labeling the $n$ bands (Note that grouping two Majorana fermions leads to one ordinary fermion), and the Majorana fermions are their own antiparticles. Then the Majorana number is given by
$\mathcal{M}=\mathrm{sgn}\Big(\mathrm{Pf}\,B(0)\Big)\mathrm{sgn}\Big(\mathrm{Pf}\,B(\pi)\Big)$.
Here $\mathrm{Pf}(X)$ is the Pfaffian of the matrix $X$.
It can be shown that the Majorana number is equivalent to the number of the edge modes \cite{Kitaev}.

The correlations of the original fermions can be expressed in terms of the Majorana-fermion correlations $\ep{a_ia_j}$. One can verify that $\ep{a_ia_i}=1$ and $\ep{a_ia_j}$ is pure imaginary if $i \neq j$. Thus, we define $\ep{a_ia_j}=iK_{ij}$ and $K_{ij}$ satisfies $K_{ij}=-K_{ji}$. For $i\neq j$,
\be
\ep{c_ic_j}&=&\frac14\Big[-(K_{2i-1,2j}+K_{2i,2j-1}) \nn\\
& &+i(K_{2i-1,2j-1}-K_{2i,2j})\Big], \\
\ep{\dc_ic_j}&=&\frac14\Big[-(K_{2i-1,2j}-K_{2i,2j-1})\nn\\
& &+i(K_{2i-1,2j-1}+K_{2i,2j})\Big].
\ee
Moreover,
\be
&&\ep{c_ic_i}=0,~\ep{\dc_ic_i}=\frac12\Big[1-K_{2i-1,2i}\Big].
\ee

\textbf{Periodic boundary condition} ---  When the boundary condition is periodic, the energy spectrum and eigenfunctions can be found analytically in momentum space by introducing the lattice Fourier transform
$c_j=\frac1{\sqrt{N}}\sum_{n} c_k e^{ikj}$ and $\dc_j=\frac1{\sqrt{N}}\sum_{n} \dc_k e^{-ikj}$. Here $k=\frac{2\pi}{N}n$ with integer $n$ in the range $-N/2+1\leq n\leq N/2$. For convenience, we take $N$ as an even number.
Then the Hamiltonian with $\mu_j=\mu_0$ for all $j$ becomes
\be
H_K&=&\sum_{k>0}\Big[\xi_k\dc_kc_k+\xi_k\dc_{-k}c_{-k}\nn\\
& &+2i\Delta_0\sin k c_{-k}c_{k}-2i\Delta_0^*\sin k\dc_k\dc_{-k}\Big]
\ee
with $\xi_k=-2w_0\cos k-\mu_0$. It can be diagonalized by the Bogoliubov transformation
\be
c_k=u_k a_k+v_k \da_{-k},\quad \dc_k=u_k \da_k+v_k^* a_{-k}.
\ee
Here we assume that $u_k$ is real and $v_k$ is complex. Then,
\be
u_k=\sqrt{\frac{E_k+\xi_k}{2E_k}},\quad v_k=ie^{i\phi}\mbox{sgn}k \sqrt{\frac{E_k-\xi_k}{2E_k}}.
\ee
Here $\phi$ is the phase of $\Delta_0=|\Delta_0|e^{i\phi}$ and sgn$k$ is the sign of $k$. We also define $E_k=\sqrt{\xi_k^2+4|\Delta_0|^2\sin^2k}$.

Then the ground state satisfying $a_k\ket{S_0}=0$ leads to the correlation functions ($a,b$ denote the sites)
\be
&&G_{ab}^{11}=\sum_k e^{ik(a-b)}u_k v_{-k},~ G_{ab}^{12}=\sum_k e^{ik(a-b)}u_k^2,\\
&&G_{ab}^{21}=\sum_k e^{-ik(a-b)}|v_k|^2,~ G_{ab}^{22}=\sum_k e^{ik(a-b)}u_kv^*_{-k}.
\ee

\textbf{Open boundary condition} --- For the open boundary case, there is no translational invariance. This makes it difficult to simplify the problem by transforming to momentum space. Instead, we will closely follow the method in Ref. \onlinecite{Lieb}, which implements a Bogoliubov transformation in real space.
The Hamiltonian can be written in the form
\be
H_K=\sum_{i,j}\Big[\dc_i A_{ij} c_j+\frac12(\dc_i B_{ij}\dc_j-c_i B_{ij}c_j)\Big].
\ee
Here $A$ is a Hermitian matrix and $B$ is an anti-symmetric matrix. For the Kitaev model with uniform parameter,
$A_{ij}=-\mu_0\delta_{ij}-w_0(\delta_{i,j+1}+\delta_{i+1,j})$ and
$B_{ij}=-\Delta_0\delta_{i,j+1}+\Delta_0\delta_{i+1,j}$.
We introduce the quasi-particle annihilation and creation operators
\be
\eta_k=\sum_i(g_{ki}c_i+h_{ki}\dc_i),~\eta_k^{\dagger}=\sum_i(g_{ki}\dc_i+h_{ki}c_i),
\ee
and define $\phi_{ki}=g_{ki}+h_{ki}$ and $\psi_{ki}=g_{ki}-h_{ki}$. By requiring that $\eta_k$ diagonalizes the Hamiltonian, we find that $\phi_{ki}$ and $\psi_{ki}$ satisfy
\be\label{eq:AB}
(A+B)_{ij}\phi_{kj}=E_k\psi_{ki},~ (A-B)_{ij}\psi_{kj}=E_k\phi_{ki}.
\ee
Here $E_k$ is the eigenvalues of the above matrices, not to be confused with the superfluid quasi-particle dispersion. Therefore, we can find $\phi_{k}$ from
$[(A-B)(A+B)]_{ij}\phi_{kj}=E_k^2\phi_{ki}$.
$\phi_{ki}$ can be chosen to be real and satisfy $\sum_k\phi_{ki}\phi_{kj}=\delta_{ij}$, and the same is true for $\psi_{ik}$. The Hamiltonian becomes
$H_K=\sum_k E_k\eta_k^{\dagger}\eta_k+\frac12(\sum_i A_{ii}-\sum_k E_k)$.

The Majorana fermion operators can be expressed as
$a_{2i-1}=\sum_k(\phi_{ki}\eta_k+\phi_{ki}\eta_k^{\dagger})$ and
$a_{2i}=i\sum_k(\psi_{ki}\eta_k^{\dagger}-\psi_{ki}\eta_k)$.
Then the correlation functions are given by
$\langle a_{2i-1}a_{2j-1}\rangle=\langle a_{2i}a_{2j}\rangle=\delta_{ij}$ and
$\langle a_{2i-1}a_{2j}\rangle=-\langle a_{2j}a_{2i-1}\rangle=i\sum_k\phi_{ki}\psi_{kj}$.
From the correlations of Majorana fermions, one can find the correlation of original fermions as
\be
G_{ij}^{11}&=&-\frac14\Big(\sum_k\phi_{ki}\psi_{kj}-\sum_k\phi_{kj}\psi_{ki}\Big),\\
G_{ij}^{12}&=&\frac14\Big(2\delta_{ij}+\sum_k\phi_{ki}\psi_{kj}+\sum_k\phi_{kj}\psi_{ki}\Big),
\ee
and similar expressions for $G_{ij}^{21}$ and $G_{ij}^{22}$.

We can also express the quasi-particles creation and annihilation operators in terms of the Majorana fermion operators as
\be \label{eq:Kitaev_Majorana}
\eta_k=\sum_i(\phi_{ki}a_{2i-1}+i\psi_{ki}a_{2i})/2,\nonumber \\
\eta_k^{\dagger}=\sum_i(\phi_{ki}a_{2i-1}-i\psi_{ki}a_{2i})/2.
\ee
The density of quasi-particles $\eta_k$ is given by
\be \label{eq:Majorana_density}
\langle\eta_k^{\dagger}\eta_k\rangle&=&\frac12+\frac i4
\sum_{ij}\Big(\phi_{ki}\psi_{kj}\langle a_{2i-1}a_{2j}\rangle
-\psi_{ki}\phi_{kj}\langle a_{2i}a_{2j-1}\rangle\Big). \nonumber \\
&&
\ee

\section{Details of SSH model}\label{app:SSH}
The initial conditions of the SSH model can also be analyzed in a similar fashion.

\textbf{Periodic boundary condition} --- Using lattice Fourier transforms
$c_{A,j}=\frac1{\sqrt{N}}\sum_{n=1}^N a_k e^{ikj}$ and $c_{B,j}=\frac1{\sqrt{N}}\sum_{n=1}^N b_k e^{ikj}$,
the Hamiltonian with $w_i=w$ and $v_i=v$ becomes
\be
H_S&=&\sum_k\psi_k^{\dagger}\Big[d_x\sigma_x+d_y\sigma_y+\mu\sigma_z\Big]\psi_k.
\ee
Here we define $\psi_k=(a_k,b_k)^T$, $d_x=(w+v\cos k)$, and $d_y=v\sin k$.
The Hamiltonian can be diagonalized by introducing
\be
&&a_k=u_k\eta_{1k}-v_k^*\eta_{2k},\quad
b_k=v_k\eta_{1k}+u_k\eta_{2k},\\
&&u_k=\sqrt{\frac{E_k+\mu}{2E_k}},\quad
v_k=\frac{d_x+id_y}{\sqrt{d_x^2+d_y^2}}\sqrt{\frac{E_k-\mu}{2E_k}}.
\ee
Here $E_k=\sqrt{d_x^2+d_y^2+\mu^2}$.
Then the Hamiltonian becomes
\be
H=\sum_k (E_k\eta_{1k}^{\dagger}\eta_{1k}-E_k\eta_{2k}^{\dagger}\eta_{2k}).
\ee

We assume that the SSH chain is half filled, so the lower band is fully filled and the upper band is empty.
The equal-time correlations are given by
\be
\langle \dc_{A,i}c_{A,j}\rangle&=&\frac1{N}\sum_k |v_k|^2 e^{-ik(i-j)},\\
\langle \dc_{B,i}c_{B,j}\rangle&=&\frac1{N}\sum_k u_k^2 e^{-ik(i-j)},\\
\langle \dc_{A,i}c_{B,j}\rangle&=&-\frac1N\sum_k u_k v_k e^{-ik(i-j)},\\
\langle \dc_{B,i}c_{A,j}\rangle&=&-\frac1N\sum_k u_k v_k^* e^{-ik(i-j)}.
\ee

\textbf{Open boundary condition} --- For the SSH model with open boundary condition, the Hamiltonian in the matrix form is
\be
&&H_{S,2i-1,2i-1}=\mu,\quad H_{S,2i,2i}=-\mu, \\
&&H_{S,2i-1,2i}=H_{2i,2i-1}=w,\\
&&H_{S,2i,2i+1}=H_{2i+1,2i}=v.
\ee
Here odd numbers label the A-sites and even numbers label the B-sites.
Then we can numerically find the $m$-th eigenvector $\phi_{im}$ corresponding to eigenvalue $E_m$, where $i=1,\cdots 2N$ labels the components. Since the Hamiltonian matrix is symmetric, we can choose all the eigenvector to be real vectors.

In terms of the quasi-particle operator $\eta_m=\sum_i\phi_{im}c_i$, the Hamiltonian becomes
\be
H_S=\sum_m E_m \eta_m^{\dagger}\eta_m.
\ee
Then it is straightforward to evaluate the equal-time correlation functions as
\be
\ep{\dc_i c_j}=\sum_{E_m<0}\phi_{im}\phi_{jm}.
\ee

\bibliographystyle{apsrev4-1}

\begin{thebibliography}{34}%
	\makeatletter
	\providecommand \@ifxundefined [1]{%
		\@ifx{#1\undefined}
	}%
	\providecommand \@ifnum [1]{%
		\ifnum #1\expandafter \@firstoftwo
		\else \expandafter \@secondoftwo
		\fi
	}%
	\providecommand \@ifx [1]{%
		\ifx #1\expandafter \@firstoftwo
		\else \expandafter \@secondoftwo
		\fi
	}%
	\providecommand \natexlab [1]{#1}%
	\providecommand \enquote  [1]{``#1''}%
	\providecommand \bibnamefont  [1]{#1}%
	\providecommand \bibfnamefont [1]{#1}%
	\providecommand \citenamefont [1]{#1}%
	\providecommand \href@noop [0]{\@secondoftwo}%
	\providecommand \href [0]{\begingroup \@sanitize@url \@href}%
	\providecommand \@href[1]{\@@startlink{#1}\@@href}%
	\providecommand \@@href[1]{\endgroup#1\@@endlink}%
	\providecommand \@sanitize@url [0]{\catcode `\\12\catcode `\$12\catcode
		`\&12\catcode `\#12\catcode `\^12\catcode `\_12\catcode `\%12\relax}%
	\providecommand \@@startlink[1]{}%
	\providecommand \@@endlink[0]{}%
	\providecommand \url  [0]{\begingroup\@sanitize@url \@url }%
	\providecommand \@url [1]{\endgroup\@href {#1}{\urlprefix }}%
	\providecommand \urlprefix  [0]{URL }%
	\providecommand \Eprint [0]{\href }%
	\providecommand \doibase [0]{http://dx.doi.org/}%
	\providecommand \selectlanguage [0]{\@gobble}%
	\providecommand \bibinfo  [0]{\@secondoftwo}%
	\providecommand \bibfield  [0]{\@secondoftwo}%
	\providecommand \translation [1]{[#1]}%
	\providecommand \BibitemOpen [0]{}%
	\providecommand \bibitemStop [0]{}%
	\providecommand \bibitemNoStop [0]{.\EOS\space}%
	\providecommand \EOS [0]{\spacefactor3000\relax}%
	\providecommand \BibitemShut  [1]{\csname bibitem#1\endcsname}%
	\let\auto@bib@innerbib\@empty
	\bibitem [{\citenamefont {Hasan}\ and\ \citenamefont {Kane}(2010)}]{Hasan10}%
	\BibitemOpen
	\bibfield  {author} {\bibinfo {author} {\bibfnamefont {M.~Z.}\ \bibnamefont
			{Hasan}}\ and\ \bibinfo {author} {\bibfnamefont {C.~L.}\ \bibnamefont
			{Kane}},\ }\href@noop {} {\bibfield  {journal} {\bibinfo  {journal} {Rev.
				Mod. Phys.}\ }\textbf {\bibinfo {volume} {82}},\ \bibinfo {pages} {3045}
		(\bibinfo {year} {2010})}\BibitemShut {NoStop}%
	\bibitem [{\citenamefont {Qi}\ and\ \citenamefont {Zhang}(2011)}]{Qi11}%
	\BibitemOpen
	\bibfield  {author} {\bibinfo {author} {\bibfnamefont {X.~L.}\ \bibnamefont
			{Qi}}\ and\ \bibinfo {author} {\bibfnamefont {S.~C.}\ \bibnamefont {Zhang}},\
	}\href@noop {} {\bibfield  {journal} {\bibinfo  {journal} {Rev. Mod. Phys.}\
	}\textbf {\bibinfo {volume} {83}},\ \bibinfo {pages} {1057} (\bibinfo {year}
	{2011})}\BibitemShut {NoStop}%
\bibitem [{\citenamefont {Bernevig}\ and\ \citenamefont
	{Hughes}(2013)}]{Bernevig_book}%
\BibitemOpen
\bibfield  {author} {\bibinfo {author} {\bibfnamefont {B.~A.}\ \bibnamefont
		{Bernevig}}\ and\ \bibinfo {author} {\bibfnamefont {T.~L.}\ \bibnamefont
		{Hughes}},\ }\href@noop {} {\emph {\bibinfo {title} {Topological insulators
			and topological superconductors}}}\ (\bibinfo  {publisher} {Princeton
	University Press},\ \bibinfo {year} {2013})\BibitemShut {NoStop}%
\bibitem [{\citenamefont {Shen}(2012)}]{ShenBook}%
\BibitemOpen
\bibfield  {author} {\bibinfo {author} {\bibfnamefont {S.~Q.}\ \bibnamefont
		{Shen}},\ }\href@noop {} {\emph {\bibinfo {title} {Topological Insulators:
			Dirac Equation in Condensed Matters}}}\ (\bibinfo  {publisher}
{Springer-Verlag},\ \bibinfo {year} {2012})\BibitemShut {NoStop}%
\bibitem [{\citenamefont {Asboth}\ \emph {et~al.}(2016)\citenamefont {Asboth},
	\citenamefont {Oroszlany},\ and\ \citenamefont {Palyi}}]{AsbothBook}%
\BibitemOpen
\bibfield  {author} {\bibinfo {author} {\bibfnamefont {J.~K.}\ \bibnamefont
		{Asboth}}, \bibinfo {author} {\bibfnamefont {L.}~\bibnamefont {Oroszlany}}, \
	and\ \bibinfo {author} {\bibfnamefont {A.}~\bibnamefont {Palyi}},\
}\href@noop {} {\emph {\bibinfo {title} {A short course on topological
		insulators: Band-structure topology and edge states in one and two
		dimensions}}}\ (\bibinfo  {publisher} {Springer},\ \bibinfo {year}
{2016})\BibitemShut {NoStop}%
\bibitem [{\citenamefont {Miyake}\ \emph {et~al.}(2013)\citenamefont {Miyake},
	\citenamefont {Siviloglou}, \citenamefont {Kennedy}, \citenamefont {Burton},\
	and\ \citenamefont {Ketterle}}]{Miyake:2013}%
\BibitemOpen
\bibfield  {author} {\bibinfo {author} {\bibfnamefont {H.}~\bibnamefont
		{Miyake}}, \bibinfo {author} {\bibfnamefont {G.~A.}\ \bibnamefont
		{Siviloglou}}, \bibinfo {author} {\bibfnamefont {C.~J.}\ \bibnamefont
		{Kennedy}}, \bibinfo {author} {\bibfnamefont {W.~C.}\ \bibnamefont {Burton}},
	\ and\ \bibinfo {author} {\bibfnamefont {W.}~\bibnamefont {Ketterle}},\
}\href {\doibase 10.1103/PhysRevLett.111.185302} {\bibfield  {journal}
{\bibinfo  {journal} {Phys. Rev. Lett.}\ }\textbf {\bibinfo {volume} {111}},\
\bibinfo {pages} {185302} (\bibinfo {year} {2013})}\BibitemShut {NoStop}%
\bibitem [{\citenamefont {Aidelsburger}\ \emph {et~al.}(2013)\citenamefont
	{Aidelsburger}, \citenamefont {Atala}, \citenamefont {Lohse}, \citenamefont
	{Barreiro}, \citenamefont {Paredes},\ and\ \citenamefont
	{Bloch}}]{Aidelsburger:2013}%
\BibitemOpen
\bibfield  {author} {\bibinfo {author} {\bibfnamefont {M.}~\bibnamefont
		{Aidelsburger}}, \bibinfo {author} {\bibfnamefont {M.}~\bibnamefont {Atala}},
	\bibinfo {author} {\bibfnamefont {M.}~\bibnamefont {Lohse}}, \bibinfo
	{author} {\bibfnamefont {J.~T.}\ \bibnamefont {Barreiro}}, \bibinfo {author}
	{\bibfnamefont {B.}~\bibnamefont {Paredes}}, \ and\ \bibinfo {author}
	{\bibfnamefont {I.}~\bibnamefont {Bloch}},\ }\href {\doibase
	10.1103/PhysRevLett.111.185301} {\bibfield  {journal} {\bibinfo  {journal}
		{Phys. Rev. Lett.}\ }\textbf {\bibinfo {volume} {111}},\ \bibinfo {pages}
	{185301} (\bibinfo {year} {2013})}\BibitemShut {NoStop}%
\bibitem [{\citenamefont {Jotzu}\ \emph {et~al.}(2014)\citenamefont {Jotzu},
	\citenamefont {Messer}, \citenamefont {Desbuquois}, \citenamefont {Lebrat},
	\citenamefont {Uehlinger}, \citenamefont {Greif},\ and\ \citenamefont
	{Esslinger}}]{Jotzu:2014}%
\BibitemOpen
\bibfield  {author} {\bibinfo {author} {\bibfnamefont {G.}~\bibnamefont
		{Jotzu}}, \bibinfo {author} {\bibfnamefont {M.}~\bibnamefont {Messer}},
	\bibinfo {author} {\bibfnamefont {R.}~\bibnamefont {Desbuquois}}, \bibinfo
	{author} {\bibfnamefont {M.}~\bibnamefont {Lebrat}}, \bibinfo {author}
	{\bibfnamefont {T.}~\bibnamefont {Uehlinger}}, \bibinfo {author}
	{\bibfnamefont {D.}~\bibnamefont {Greif}}, \ and\ \bibinfo {author}
	{\bibfnamefont {T.}~\bibnamefont {Esslinger}},\ }\href
{http://dx.doi.org/10.1038/nature13915} {\bibfield  {journal} {\bibinfo
		{journal} {Nature}\ }\textbf {\bibinfo {volume} {515}},\ \bibinfo {pages}
	{237} (\bibinfo {year} {2014})}\BibitemShut {NoStop}%
\bibitem [{\citenamefont {Lu}\ \emph {et~al.}(2014)\citenamefont {Lu},
	\citenamefont {Joannopoulos},\ and\ \citenamefont {Soljacic}}]{Lu14}%
\BibitemOpen
\bibfield  {author} {\bibinfo {author} {\bibfnamefont {L.}~\bibnamefont
		{Lu}}, \bibinfo {author} {\bibfnamefont {J.~D.}\ \bibnamefont
		{Joannopoulos}}, \ and\ \bibinfo {author} {\bibfnamefont {M.}~\bibnamefont
		{Soljacic}},\ }\href@noop {} {\bibfield  {journal} {\bibinfo  {journal} {Nat.
			Photonics}\ }\textbf {\bibinfo {volume} {8}},\ \bibinfo {pages} {821}
	(\bibinfo {year} {2014})}\BibitemShut {NoStop}%
\bibitem [{\citenamefont {Yang}\ \emph {et~al.}(2015)\citenamefont {Yang},
	\citenamefont {Gao}, \citenamefont {Shi}, \citenamefont {Lin}, \citenamefont
	{Gao}, \citenamefont {Chong},\ and\ \citenamefont {Zhang}}]{Yang15}%
\BibitemOpen
\bibfield  {author} {\bibinfo {author} {\bibfnamefont {Z.}~\bibnamefont
		{Yang}}, \bibinfo {author} {\bibfnamefont {F.}~\bibnamefont {Gao}}, \bibinfo
	{author} {\bibfnamefont {X.}~\bibnamefont {Shi}}, \bibinfo {author}
	{\bibfnamefont {X.}~\bibnamefont {Lin}}, \bibinfo {author} {\bibfnamefont
		{Z.}~\bibnamefont {Gao}}, \bibinfo {author} {\bibfnamefont {Y.}~\bibnamefont
		{Chong}}, \ and\ \bibinfo {author} {\bibfnamefont {B.}~\bibnamefont
		{Zhang}},\ }\href@noop {} {\bibfield  {journal} {\bibinfo  {journal} {Phys.
			Rev. Lett.}\ }\textbf {\bibinfo {volume} {114}},\ \bibinfo {pages} {114301}
	(\bibinfo {year} {2015})}\BibitemShut {NoStop}%
\bibitem [{\citenamefont {Sacramento}(2014)}]{Sacramento14}%
\BibitemOpen
\bibfield  {author} {\bibinfo {author} {\bibfnamefont {P.~D.}\ \bibnamefont
		{Sacramento}},\ }\href@noop {} {\bibfield  {journal} {\bibinfo  {journal}
		{Phys. Rev. E}\ }\textbf {\bibinfo {volume} {90}},\ \bibinfo {pages} {032138}
	(\bibinfo {year} {2014})}\BibitemShut {NoStop}%
\bibitem [{\citenamefont {Caio}\ \emph {et~al.}(2015)\citenamefont {Caio},
	\citenamefont {Cooper},\ and\ \citenamefont {Bhaseen}}]{Caio15}%
\BibitemOpen
\bibfield  {author} {\bibinfo {author} {\bibfnamefont {M.~D.}\ \bibnamefont
		{Caio}}, \bibinfo {author} {\bibfnamefont {N.~R.}\ \bibnamefont {Cooper}}, \
	and\ \bibinfo {author} {\bibfnamefont {M.~J.}\ \bibnamefont {Bhaseen}},\
}\href@noop {} {\bibfield  {journal} {\bibinfo  {journal} {Phys. Rev. Lett.}\
}\textbf {\bibinfo {volume} {115}},\ \bibinfo {pages} {236403} (\bibinfo
{year} {2015})}\BibitemShut {NoStop}%
\bibitem [{\citenamefont {Wang}\ \emph {et~al.}(2015)\citenamefont {Wang},
	\citenamefont {Schmitt},\ and\ \citenamefont {Kehrein}}]{Wang15}%
\BibitemOpen
\bibfield  {author} {\bibinfo {author} {\bibfnamefont {P.}~\bibnamefont
		{Wang}}, \bibinfo {author} {\bibfnamefont {M.}~\bibnamefont {Schmitt}}, \
	and\ \bibinfo {author} {\bibfnamefont {S.}~\bibnamefont {Kehrein}},\
}\href@noop {} {\enquote {\bibinfo {title} {Universal nonanalytic behavior of
		the hall conductance in a chern insulator at the topologically driven
		nonequilibrium phase transition},}\ } (\bibinfo {year} {2015}),\ \bibinfo
{note} {arXiv: 1511.05255}\BibitemShut {NoStop}%
\bibitem [{\citenamefont {Zeng}\ \emph {et~al.}(2015)\citenamefont {Zeng},
	\citenamefont {Hamma},\ and\ \citenamefont {Fan}}]{Zeng15}%
\BibitemOpen
\bibfield  {author} {\bibinfo {author} {\bibfnamefont {Y.}~\bibnamefont
		{Zeng}}, \bibinfo {author} {\bibfnamefont {A.}~\bibnamefont {Hamma}}, \ and\
	\bibinfo {author} {\bibfnamefont {H.}~\bibnamefont {Fan}},\ }\href@noop {}
{\enquote {\bibinfo {title} {Thermalization of topological entropy after a
			quantum quench},}\ } (\bibinfo {year} {2015}),\ \bibinfo {note} {arXiv:
	1509.08613}\BibitemShut {NoStop}%
\bibitem [{\citenamefont {Sacramento}(2016)}]{Sacramento16}%
\BibitemOpen
\bibfield  {author} {\bibinfo {author} {\bibfnamefont {P.~D.}\ \bibnamefont
		{Sacramento}},\ }\href@noop {} {\enquote {\bibinfo {title} {Edge mode
			dynamics of quenched topological wires},}\ } (\bibinfo {year} {2016}),\
\bibinfo {note} {arXiv: 1601.05476}\BibitemShut {NoStop}%
\bibitem [{\citenamefont {Grushin}\ \emph {et~al.}(2015)\citenamefont
	{Grushin}, \citenamefont {Roy},\ and\ \citenamefont {Haque}}]{Grushin15}%
\BibitemOpen
\bibfield  {author} {\bibinfo {author} {\bibfnamefont {A.~G.}\ \bibnamefont
		{Grushin}}, \bibinfo {author} {\bibfnamefont {S.}~\bibnamefont {Roy}}, \ and\
	\bibinfo {author} {\bibfnamefont {M.}~\bibnamefont {Haque}},\ }\href@noop {}
{\enquote {\bibinfo {title} {Response of fermions in chern bands to spatially
			local quenches},}\ } (\bibinfo {year} {2015}),\ \bibinfo {note} {arXiv:
	1508.04778}\BibitemShut {NoStop}%
\bibitem [{\citenamefont {Kitaev}(2001)}]{Kitaev}%
\BibitemOpen
\bibfield  {author} {\bibinfo {author} {\bibfnamefont {A.~Y.}\ \bibnamefont
		{Kitaev}},\ }\href@noop {} {\bibfield  {journal} {\bibinfo  {journal} {Phys.
			-Usp.}\ }\textbf {\bibinfo {volume} {44}},\ \bibinfo {pages} {131} (\bibinfo
	{year} {2001})}\BibitemShut {NoStop}%
\bibitem [{\citenamefont {Su}\ \emph {et~al.}(1979)\citenamefont {Su},
	\citenamefont {Schrieffer},\ and\ \citenamefont {Heeger}}]{SSH79}%
\BibitemOpen
\bibfield  {author} {\bibinfo {author} {\bibfnamefont {W.~P.}\ \bibnamefont
		{Su}}, \bibinfo {author} {\bibfnamefont {J.~R.}\ \bibnamefont {Schrieffer}},
	\ and\ \bibinfo {author} {\bibfnamefont {A.~J.}\ \bibnamefont {Heeger}},\
}\href@noop {} {\bibfield  {journal} {\bibinfo  {journal} {Phys. Rev. Lett.}\
}\textbf {\bibinfo {volume} {42}},\ \bibinfo {pages} {1698} (\bibinfo {year}
{1979})}\BibitemShut {NoStop}%
\bibitem [{\citenamefont {Atala}\ \emph {et~al.}(2013)\citenamefont {Atala},
	\citenamefont {Aidelsburger}, \citenamefont {Barreiro}, \citenamefont
	{Abanin}, \citenamefont {Kitagawa}, \citenamefont {Demler},\ and\
	\citenamefont {Bloch}}]{Atala13}%
\BibitemOpen
\bibfield  {author} {\bibinfo {author} {\bibfnamefont {M.}~\bibnamefont
		{Atala}}, \bibinfo {author} {\bibfnamefont {M.}~\bibnamefont {Aidelsburger}},
	\bibinfo {author} {\bibfnamefont {J.~T.}\ \bibnamefont {Barreiro}}, \bibinfo
	{author} {\bibfnamefont {D.}~\bibnamefont {Abanin}}, \bibinfo {author}
	{\bibfnamefont {T.}~\bibnamefont {Kitagawa}}, \bibinfo {author}
	{\bibfnamefont {E.}~\bibnamefont {Demler}}, \ and\ \bibinfo {author}
	{\bibfnamefont {I.}~\bibnamefont {Bloch}},\ }\href@noop {} {\bibfield
	{journal} {\bibinfo  {journal} {Nat. Phys.}\ }\textbf {\bibinfo {volume}
		{9}},\ \bibinfo {pages} {795} (\bibinfo {year} {2013})}\BibitemShut {NoStop}%
\bibitem [{\citenamefont {Ryu}\ \emph {et~al.}(2010)\citenamefont {Ryu},
	\citenamefont {Schnyder}, \citenamefont {Furusaki},\ and\ \citenamefont
	{Ludwig}}]{Ryu10}%
\BibitemOpen
\bibfield  {author} {\bibinfo {author} {\bibfnamefont {S.}~\bibnamefont
		{Ryu}}, \bibinfo {author} {\bibfnamefont {A.~P.}\ \bibnamefont {Schnyder}},
	\bibinfo {author} {\bibfnamefont {A.}~\bibnamefont {Furusaki}}, \ and\
	\bibinfo {author} {\bibfnamefont {A.~W.~W.}\ \bibnamefont {Ludwig}},\
}\href@noop {} {\bibfield  {journal} {\bibinfo  {journal} {New J. Phys.}\
}\textbf {\bibinfo {volume} {12}},\ \bibinfo {pages} {065010} (\bibinfo
{year} {2010})}\BibitemShut {NoStop}%
\bibitem [{\citenamefont {Lai}\ and\ \citenamefont {Chien}(2016)}]{Lai16}%
\BibitemOpen
\bibfield  {author} {\bibinfo {author} {\bibfnamefont {C.~Y.}\ \bibnamefont
		{Lai}}\ and\ \bibinfo {author} {\bibfnamefont {C.~C.}\ \bibnamefont
		{Chien}},\ }\href@noop {} {\bibfield  {journal} {\bibinfo  {journal} {Phys.
			Rev. Applied.}\ }\textbf {\bibinfo {volume} {5}},\ \bibinfo {pages} {034001}
	(\bibinfo {year} {2016})}\BibitemShut {NoStop}%
\bibitem [{\citenamefont {Cornean}\ \emph {et~al.}(2013)\citenamefont
	{Cornean}, \citenamefont {Jensen},\ and\ \citenamefont {Nenciu}}]{Cornean13}%
\BibitemOpen
\bibfield  {author} {\bibinfo {author} {\bibfnamefont {H.~D.}\ \bibnamefont
		{Cornean}}, \bibinfo {author} {\bibfnamefont {A.}~\bibnamefont {Jensen}}, \
	and\ \bibinfo {author} {\bibfnamefont {G.}~\bibnamefont {Nenciu}},\
}\href@noop {} {\bibfield  {journal} {\bibinfo  {journal} {Ann. Henri
		Poincar{\'e}}\ }\textbf {\bibinfo {volume} {15}},\ \bibinfo {pages} {1919}
(\bibinfo {year} {2013})}\BibitemShut {NoStop}%
\bibitem [{\citenamefont {Metcalf}\ \emph
	{et~al.}(2016{\natexlab{a}})\citenamefont {Metcalf}, \citenamefont {Lai},\
	and\ \citenamefont {Chien}}]{MekenaHyst}%
\BibitemOpen
\bibfield  {author} {\bibinfo {author} {\bibfnamefont {M.}~\bibnamefont
		{Metcalf}}, \bibinfo {author} {\bibfnamefont {C.~Y.}\ \bibnamefont {Lai}}, \
	and\ \bibinfo {author} {\bibfnamefont {C.~C.}\ \bibnamefont {Chien}},\
}\href@noop {} {\bibfield  {journal} {\bibinfo  {journal} {Phys. Rev. A}\
}\textbf {\bibinfo {volume} {93}},\ \bibinfo {pages} {053617} (\bibinfo
{year} {2016}{\natexlab{a}})}\BibitemShut {NoStop}%
\bibitem [{\citenamefont {Fagotti}()}]{Fagotti16}%
\BibitemOpen
\bibfield  {author} {\bibinfo {author} {\bibfnamefont {M.}~\bibnamefont
		{Fagotti}},\ }\href@noop {} {\enquote {\bibinfo {title} {Control of global
			properties in a closed many-body quantum system by means of a local
			switch},}\ }\bibinfo {note} {ArXiv: 1508.04401}\BibitemShut {NoStop}%
\bibitem [{\citenamefont {Li}\ \emph {et~al.}(2014)\citenamefont {Li},
	\citenamefont {Xu},\ and\ \citenamefont {Chen}}]{Li14}%
\BibitemOpen
\bibfield  {author} {\bibinfo {author} {\bibfnamefont {L.}~\bibnamefont
		{Li}}, \bibinfo {author} {\bibfnamefont {Z.}~\bibnamefont {Xu}}, \ and\
	\bibinfo {author} {\bibfnamefont {S.}~\bibnamefont {Chen}},\ }\href@noop {}
{\bibfield  {journal} {\bibinfo  {journal} {Phys. Rev. B}\ }\textbf {\bibinfo
		{volume} {89}},\ \bibinfo {pages} {085111} (\bibinfo {year}
	{2014})}\BibitemShut {NoStop}%
\bibitem [{\citenamefont {Ashcroft}\ and\ \citenamefont
	{Mermin}(1976)}]{AshcroftBook}%
\BibitemOpen
\bibfield  {author} {\bibinfo {author} {\bibfnamefont {N.~W.}\ \bibnamefont
		{Ashcroft}}\ and\ \bibinfo {author} {\bibfnamefont {N.~D.}\ \bibnamefont
		{Mermin}},\ }\href@noop {} {\emph {\bibinfo {title} {Solid state physics}}}\
(\bibinfo  {publisher} {Thomson Learning},\ \bibinfo {address} {Stamford,
	USA},\ \bibinfo {year} {1976})\BibitemShut {NoStop}%
\bibitem [{\citenamefont {Metcalf}\ \emph
	{et~al.}(2016{\natexlab{b}})\citenamefont {Metcalf}, \citenamefont {Chern},
	\citenamefont {Di~Ventra},\ and\ \citenamefont {Chien}}]{Mekena16}%
\BibitemOpen
\bibfield  {author} {\bibinfo {author} {\bibfnamefont {M.}~\bibnamefont
		{Metcalf}}, \bibinfo {author} {\bibfnamefont {G.~W.}\ \bibnamefont {Chern}},
	\bibinfo {author} {\bibfnamefont {M.}~\bibnamefont {Di~Ventra}}, \ and\
	\bibinfo {author} {\bibfnamefont {C.~C.}\ \bibnamefont {Chien}},\ }\href@noop
{} {\bibfield  {journal} {\bibinfo  {journal} {J. Phys. B: At. Mol. Opt.
			Phys.}\ }\textbf {\bibinfo {volume} {49}},\ \bibinfo {pages} {075301}
	(\bibinfo {year} {2016}{\natexlab{b}})}\BibitemShut {NoStop}%
\bibitem [{\citenamefont {Mourik}\ \emph {et~al.}(2012)\citenamefont {Mourik},
	\citenamefont {Zuo}, \citenamefont {Frolov}, \citenamefont {Plissard},
	\citenamefont {Bakkers},\ and\ \citenamefont {Kouwenhoven}}]{Mourik12}%
\BibitemOpen
\bibfield  {author} {\bibinfo {author} {\bibfnamefont {V.}~\bibnamefont
		{Mourik}}, \bibinfo {author} {\bibfnamefont {K.}~\bibnamefont {Zuo}},
	\bibinfo {author} {\bibfnamefont {S.~M.}\ \bibnamefont {Frolov}}, \bibinfo
	{author} {\bibfnamefont {S.~R.}\ \bibnamefont {Plissard}}, \bibinfo {author}
	{\bibfnamefont {E.~P. A.~M.}\ \bibnamefont {Bakkers}}, \ and\ \bibinfo
	{author} {\bibfnamefont {L.~P.}\ \bibnamefont {Kouwenhoven}},\ }\href@noop {}
{\bibfield  {journal} {\bibinfo  {journal} {Science}\ }\textbf {\bibinfo
		{volume} {336}},\ \bibinfo {pages} {1003} (\bibinfo {year}
	{2012})}\BibitemShut {NoStop}%
\bibitem [{\citenamefont {Nadj-Perge}\ \emph {et~al.}(2014)\citenamefont
	{Nadj-Perge}, \citenamefont {Drozdov}, \citenamefont {Li}, \citenamefont
	{Chen}, \citenamefont {Jeon}, \citenamefont {Seo}, \citenamefont {MacDonald},
	\citenamefont {Bernevig},\ and\ \citenamefont {Yazdani}}]{Perge14}%
\BibitemOpen
\bibfield  {author} {\bibinfo {author} {\bibfnamefont {S.}~\bibnamefont
		{Nadj-Perge}}, \bibinfo {author} {\bibfnamefont {I.~K.}\ \bibnamefont
		{Drozdov}}, \bibinfo {author} {\bibfnamefont {J.}~\bibnamefont {Li}},
	\bibinfo {author} {\bibfnamefont {H.}~\bibnamefont {Chen}}, \bibinfo {author}
	{\bibfnamefont {S.}~\bibnamefont {Jeon}}, \bibinfo {author} {\bibfnamefont
		{J.}~\bibnamefont {Seo}}, \bibinfo {author} {\bibfnamefont {A.~H.}\
		\bibnamefont {MacDonald}}, \bibinfo {author} {\bibfnamefont {B.~A.}\
		\bibnamefont {Bernevig}}, \ and\ \bibinfo {author} {\bibfnamefont
		{A.}~\bibnamefont {Yazdani}},\ }\href@noop {} {\bibfield  {journal} {\bibinfo
		{journal} {Science}\ }\textbf {\bibinfo {volume} {346}},\ \bibinfo {pages}
	{602} (\bibinfo {year} {2014})}\BibitemShut {NoStop}%
\bibitem [{\citenamefont {Henderson}\ \emph {et~al.}(2009)\citenamefont
	{Henderson}, \citenamefont {Ryu}, \citenamefont {MacCormick},\ and\
	\citenamefont {Boshier}}]{Henderson09}%
\BibitemOpen
\bibfield  {author} {\bibinfo {author} {\bibfnamefont {K.}~\bibnamefont
		{Henderson}}, \bibinfo {author} {\bibfnamefont {C.}~\bibnamefont {Ryu}},
	\bibinfo {author} {\bibfnamefont {C.}~\bibnamefont {MacCormick}}, \ and\
	\bibinfo {author} {\bibfnamefont {M.~G.}\ \bibnamefont {Boshier}},\
}\href@noop {} {\bibfield  {journal} {\bibinfo  {journal} {New J. Phys.}\
}\textbf {\bibinfo {volume} {11}},\ \bibinfo {pages} {043030} (\bibinfo
{year} {2009})}\BibitemShut {NoStop}%
\bibitem [{\citenamefont {Chien}\ \emph {et~al.}(2016)\citenamefont {Chien},
	\citenamefont {Peotta},\ and\ \citenamefont {Di~Ventra}}]{ChienNP}%
\BibitemOpen
\bibfield  {author} {\bibinfo {author} {\bibfnamefont {C.~C.}\ \bibnamefont
		{Chien}}, \bibinfo {author} {\bibfnamefont {S.}~\bibnamefont {Peotta}}, \
	and\ \bibinfo {author} {\bibfnamefont {M.}~\bibnamefont {Di~Ventra}},\
}\href@noop {} {\bibfield  {journal} {\bibinfo  {journal} {Nat. Phys.}\
}\textbf {\bibinfo {volume} {11}},\ \bibinfo {pages} {998} (\bibinfo {year}
{2016})}\BibitemShut {NoStop}%
\bibitem [{\citenamefont {Bakr}\ \emph {et~al.}(2009)\citenamefont {Bakr},
	\citenamefont {Gillen}, \citenamefont {Peng}, \citenamefont {Folling},\ and\
	\citenamefont {Greiner}}]{Bakr09}%
\BibitemOpen
\bibfield  {author} {\bibinfo {author} {\bibfnamefont {W.~S.}\ \bibnamefont
		{Bakr}}, \bibinfo {author} {\bibfnamefont {J.~I.}\ \bibnamefont {Gillen}},
	\bibinfo {author} {\bibfnamefont {A.}~\bibnamefont {Peng}}, \bibinfo {author}
	{\bibfnamefont {S.}~\bibnamefont {Folling}}, \ and\ \bibinfo {author}
	{\bibfnamefont {M.}~\bibnamefont {Greiner}},\ }\href@noop {} {\bibfield
	{journal} {\bibinfo  {journal} {Nature}\ }\textbf {\bibinfo {volume} {462}},\
	\bibinfo {pages} {74} (\bibinfo {year} {2009})}\BibitemShut {NoStop}%
\bibitem [{\citenamefont {Fetter}\ and\ \citenamefont
	{Walecka}(2003)}]{Walecka}%
\BibitemOpen
\bibfield  {author} {\bibinfo {author} {\bibfnamefont {A.~L.}\ \bibnamefont
		{Fetter}}\ and\ \bibinfo {author} {\bibfnamefont {J.~D.}\ \bibnamefont
		{Walecka}},\ }\href@noop {} {\emph {\bibinfo {title} {Quantum Theory of
			Many-Particle Systems}}}\ (\bibinfo  {publisher} {Dover Publications},\
\bibinfo {year} {2003})\BibitemShut {NoStop}%
\bibitem [{\citenamefont {E.~Lieb}(1961)}]{Lieb}%
\BibitemOpen
\bibfield  {author} {\bibinfo {author} {\bibfnamefont {D.~M.}\ \bibnamefont
		{E.~Lieb}, \bibfnamefont {T.~Schultz}},\ }\href@noop {} {\bibfield  {journal}
	{\bibinfo  {journal} {Ann. Phys.}\ }\textbf {\bibinfo {volume} {16}},\
	\bibinfo {pages} {407} (\bibinfo {year} {1961})}\BibitemShut {NoStop}%
\end{thebibliography}
%

\end{document}